\documentclass[structabstract]{aa}
\usepackage{graphicx}
\usepackage{txfonts}
\usepackage{natbib}
\usepackage{color}

\begin{document}
\title{The evolutionary state of Miras with changing pulsation
  periods\thanks{Based on observations made with the Mercator Telescope,
  operated on the island of La Palma by the Flemish Community, at the Spanish
  Observatorio del Roque de los Muchachos of the Instituto de Astrof\'{i}sica
  de Canarias, and with the Swiss 1.2\,m Euler telescope at La Silla, Chile.
  The spectra are only available in electronic form at the CDS via anonymous
  ftp to cdsarc.u-strasbg.fr (130.79.128.5) or via
  http://cdsweb.u-strasbg.fr/cgi-bin/qcat?J/A+A/}}
\author{Stefan Uttenthaler\inst{1,2}
  \and
  Koen Van Stiphout\inst{1,3}
  \and
  Kevin Voet\inst{1}
  \and
  Hans Van Winckel\inst{1}
  \and
  Sophie Van Eck\inst{4}
  \and
  Alain Jorissen\inst{4}
  \and
  Franz Kerschbaum\inst{2}
  \and
  Gert Raskin\inst{1}
  \and
  Saskia Prins\inst{1}
  \and
  Wim Pessemier\inst{1}
  \and
  Christoffel Waelkens\inst{1}
  \and
  Yves Fr\'{e}mat\inst{5}
  \and
  Herman Hensberge\inst{5}
  \and
  Louis Dumortier\inst{5}
  \and
  Holger Lehmann\inst{6}
}

\institute{Instituut voor Sterrenkunde, University of Leuven,
  Celestijnenlaan 200D, 3001 Leuven, Belgium\\
  \email{stefan@ster.kuleuven.be}
  \and
  University of Vienna, Department of Astronomy,
  T{\"u}rken\-schanz\-stra\ss e 17, A-1180 Vienna, Austria
  \and
  Instituut voor Kern- en Stralingsfysica, University of Leuven,
  Celestijnenlaan 200D, 3001 Leuven, Belgium
  \and
  Institut d'Astronomie et d'Astrophysique, Universit\'{e} Libre de Bruxelles,
  CP226, Boulevard de Triomphe, 1050 Brussels, Belgium
  \and
  Royal Observatory of Belgium, Ringlaan 3, 1180 Brussels, Belgium
  \and
  Th\"uringer Landessternwarte Tautenburg, Sternwarte 5, 07778 Tautenburg,
  Germany
}

\date{Received 7 January, 2011; accepted 20 April, 2011}

% \abstract{}{}{}{}{} 
% 5 {} token are mandatory
 
\abstract
    {
      Miras are long-period variables thought to be in the asymptotic giant
      branch (AGB) phase of evolution.
%      Most of them have pulsation periods of the order of a few hundred days,
%      which are stable over decades to centuries.
      In about one percent of known Miras, the pulsation period is changing.
%      Three types of secular change are distinguished: meandering, continuous,
%      and sudden change.
      It has been speculated that this changing period is the consequence of a
      recent thermal pulse in these stars.
%      ,which is a powerful ignition of the otherwise dormant helium-burning
%      shell.
    }
    {
      We aim to clarify the evolutionary state of these stars, and to determine
      in particular whether or not they are in the thermally-pulsing (TP-)AGB
      phase.
    }
    {
      One important piece of information that has been neglected so far when
      determining the evolutionary state is the presence of the radio-active
      s-process element technetium (Tc). We obtained high-resolution, high
      signal-to-noise-ratio optical spectra of a dozen prominent Mira variables
      with changing pulsation period to search for this indicator of TPs and
      dredge-up. We also use the spectra to measure lithium (Li) abundances.
      Furthermore, we establish the evolutionary states of our sample stars by
      means of their present-day periods and luminosities.
    }
   {
     Among the twelve sample stars observed in this programme, five were found
     to show absorption lines of Tc. \object{BH~Cru} is found to be a
     carbon-star, its period increase in the past decades possibly having
     stopped by now. We report a possible switch in the pulsation mode of
     \object{T~UMi} from Mira-like to semi-regular variability in the past two
     years. \object{R~Nor}, on the other hand, is probably a fairly massive AGB
     star, which could be true for all meandering Miras. Finally, we assign
     \object{RU~Vul} to the metal-poor thick disk with properties very similar
     to the short-period, metal-poor Miras.
   }
   {We conclude that there is no clear correlation between period change class
     and Tc presence. The stars that are most likely to have experienced a
     recent TP are BH~Cru and \object{R~Hya}, although their rates of period
     change are quite different.}


   \maketitle

%________________________________________________________________

\section{Introduction}

Mira-type variables are red giant stars that exhibit long-period variability
with large amplitude ($\Delta V > 2\fm5$). The pulsation period of most
Mira variables is stable over long time spans (i.e.\ of between decades and
centuries), but there are  exceptions. \citet{ZB02} define three classes of
secular period change in Mira variables:
\begin{description}
\item[{\bf Continuous change:}] A continuous decrease or increase in the
  period over a century or more, with no indication for epochs of a stable
  period. The change in period is in all cases large, 15\% or more over 100
  years. In the following, we refer to this class as CCh.
\item[{\bf Sudden change:}] After a long phase of stability, the period
  suddenly begins to either decrease or increase at a large rate. The total
  change in period is similar to that of the CCh class, but the rate is ten
  times faster because the same change is achieved within a decade rather than
  within a century (SCh in the following, not to be confused with the SC
  spectral type!).
\item[{\bf Meandering change:}] A number of long-period Miras show evidence of
  meandering or fluctuating periods. The periods change by 10\% over several
  decades, followed by a return to the previous period. The rate of change is
  comparable to that of the CCh Miras, but the total change is somewhat smaller
  than that of the other two classes. Most Miras in this class have periods in
  excess of 400\,d (MCh in the following).
\end{description}

\citet{TMW05} study the period evolution of 547 Mira variables, of which 57 are
found to have a period change at the $2 \sigma$ confidence level, and eight
stars at the level of $6 \sigma$. Their catalogue is the current reference for
observed secular evolution in Miras in the solar neighbourhood.

From the point of view of stellar evolution, Mira variables are thought to be
in the asymptotic giant branch (AGB) phase. The hydrogen-burning shell is
the main energy source during most of the time on the AGB, which is
quasi-periodically interrupted by powerful ignitions of the helium-burning
shell \citep{SH65}. These ignitions are called helium-shell flashes or thermal
pulses (TPs), hence this phase is also referred to as the TP-AGB. A TP has a
large impact on the structure of the whole star: the luminosity, temperature,
radius, and thus pulsation period of the outer envelope, are predicted to vary
strongly within centuries. For a star with 2\,$M_{\odot}$ in the middle of
its TP-AGB phase, the TP itself (convective zone in the inter-shell region)
lasts about 300 -- 400 years. The evolution during the TP is governed by the
thermal timescale of the envelope, and the evolution is faster in both later
TPs and more massive stars. Several hundred years after the onset of the TP,
surface abundance changes can occur as the third dredge-up (3DUP) begins to
operate. This mixing event may take place after a TP when the convective
envelope deepens to reach layers that have previously undergone nuclear
processing \citep{Busso99}. The onset of 3DUP is predicted to coincide
with a temporary luminosity maximum. Once the convective envelope reaches
layers with processed material, it should become visible on the stellar surface
very quickly because the turn-over time of the convective envelope is of the
order of only a few years.

The first suggestion that the observed period change in a few Mira stars could
be related to a recent TP dates back to \citet{Wood75}. \citet{WZ81} present
evolutionary models of the TP-AGB and attempt to derive constraints on the core
mass and luminosity of three Miras (\object{R~Hya}, \object{R~Aql}, and
\object{W~Dra}) with a well-known, long-term period change. The luminosity
evolution inferred from the observed period change of the stars is reproduced
well by the predicted luminosity evolution of the models. This gives some
support to the suggestion that a TP is the cause of the period change. The
fraction of Miras with large period changes \citep[$\sim$1.6\%;][]{TMW05}
agrees well with the expected ratio of pulse duration to inter-pulse time of
roughly 1 -- 2\%. Hence, from a statistical point of view, we expect that at
any given time at least 1\% of the Miras is undergoing a TP. Whether the ones
that are observed to show a large period change are also the ones that
currently undergo a TP is a different question, which we aim to address here.

There are, however, also alternative explanations of the observed period
changes, which do not require a recent TP. \citet{YT96} and \citet{LW05}
identify a feedback mechanism between the pulsation and the stellar entropy
structure, which can lead to mode switching. On the other hand, \citet{Zij04}
propose that a possibly chaotic feedback between molecular opacities, pulsation
amplitude, and period can cause an unstable period. This is supposed to be most
important for stars with a C/O ratio very close to unity (SC and CS spectral
types), as small changes in the temperature in the atmosphere can cause large
changes in the molecular abundances, and hence the opacities. This alternative
explanation might be applicable to the period change in the MCh group, as their
periods change back and forth on a much shorter timescale than can be explained
by a TP.

A piece of information that has not yet been taken into account in this
discussion is the presence of the radio-active element technetium (Tc) in the
atmospheres of these stars. Tc is a product of the slow neutron-capture process
(s-process) that takes place in deep layers of AGB stars. The s-process
products are subsequently mixed to the surface by a 3DUP event. The
longest-lived Tc isotope produced by the s-process is $^{99}$Tc with a half-life
time of 210\,000 years, which is short compared to stellar evolution timescales
before the AGB.  If observed on the surface of a star, Tc is an indisputable
sign of an ongoing s-process and 3DUP in that star, hence indicates that the
star is indeed in the TP-AGB phase of evolution. Furthermore, 3DUP mixes up
carbon, the product of He-burning, which can change the star from an
oxygen-rich M-type to a carbon-rich C-type star.

In previous surveys of Tc in red giant stars \citep[e.g.][]{Lit87}, only a few
Miras with changing pulsation periods were investigated, and for those that
had been observed the presence of Tc was often unclear. The presence of Tc
would be support for the hypothesis of a recent TP that might be the cause of
the observed period change. We note, however, that the absence of Tc does
{\em not} exclude that TPs are going on in a given star. Nevertheless, the
absence of Tc places a constraint on the strength of the TP, because only the
strongest TPs on the upper AGB will be followed by a 3DUP event.% This
%constraint is of importance if one tries to simulate these stars with detailed
%AGB evolutionary models that self-consistently predict the occurrence of 3DUP.

\smallskip
In addition to Tc, the light element lithium (Li) is an important diagnostic
tool for stellar evolution studies, in particular for internal mixing
processes. It is very fragile because it is burnt to helium by proton captures
at temperatures exceeding $3\times10^6$\,K. Standard stellar evolution predicts
that Li should diminish during the evolution to the giant phase, and indeed red
giants are generally observed to have a low Li abundance compared to main
sequence stars. However, standard stellar evolution also predicts that in high
mass AGB stars ($M \gtrsim 4 M_{\odot}$), Li may be produced in large amounts.
In these stars, the bottom of the convective envelope extends to the H-burning
shell \citep[also called hot bottom burning, or HBB;][]{SB92}, which activates
the Cameron-Fowler mechanism \citep{CF71} producing Li. Moreover, these stars
do not become C-rich by means of the dredge-up of C, because the CN cycle also
converts $^{12}$C to $^{14}$N. Li surface abundances of up to
$\log\epsilon({\rm Li}) = 4.5$ (where
$\log\epsilon({\rm Li}) = \log N ({\rm Li})/ N ({\rm H})+12$) are predicted
\citep[e.g.][]{SB92,Amanda}. Luminous, Li-rich stars have been observed in both
the Magellanic Clouds \citep{Smith95} and the Galaxy \citep{GH07}. Hence, the
observation of a luminous O- and Li-rich star infers a fairly high mass for
that star.

Lower-mass AGB stars can also be somewhat enriched in Li. The layer of
radiative energy transport, which usually prevents the operation of the
Cameron-Fowler mechanism in these stars, is naturally overcome during 3DUP
events and can lead to Li enhancement at the surface up to
$\log\epsilon(\rm{Li}) \lesssim 1.8$ \citep{Kar10}. In addition to this,
non-convective extra-mixing processes (e.g.\ thermohaline mixing, magnetic
buoyancy) may provide transport of matter across the radiative barrier.
Observational evidence along these lines is presented by \citet{Van07},
\citet{Utt07b}, and \citet{UL10}. However, measurements of the Li abundances of
O-rich, M-type AGB stars remain scarce.

For the present work, we selected several prominent representatives of the
three period change classes described above and observed their optical
spectra at high resolution. With these data, we can safely establish the
evolutionary state of the Tc-rich stars on the TP-AGB. A central aim of the
paper is to check whether the presence of Tc is correlated with the different
period change classes. The spectra also provided us with the opportunity to
measure the abundance of Li. In addition, we determined the present-day
pulsation periods of the stars from AAVSO observations, to determine whether the
period change has continued in recent years. To construct an HR diagram of the
stars, we calculated their luminosities with a period-magnitude relation.

\smallskip
The structure of the paper is as follows. In Sect.~\ref{sectObs}, the target
selection and observations are described; Sect.~\ref{AnalRes} is devoted to the
analysis of the spectra and the results thereof; the results are discussed in
Sect.~\ref{Disc}, and finally Sect.~\ref{Concl} presents our conclusions.

\section{Target selection and observations}\label{sectObs}

The targets of our observations were selected from the stars discussed in
\citet{WZ81} and \citet{ZB02}. They are prominent representatives of the
period change classes introduced by \citet{ZB02}, and most are also discussed
in detail in \citet{TMW05}. In total, twelve stars were selected for the
observations.

The targets that are most readily observable from the northern hemisphere were
observed with the Hermes spectrograph \citep{Ras2010}, which is mounted on the
1.2\,m Mercator telescope at the Roque de los Muchachos observatory on the
island of La Palma, Spain. The two targets best observable from the southern
hemisphere were observed with the Coralie spectrograph, which is mounted on the
1.2\,m Leonhard Euler telescope at La Silla observatory, Chile. Both Hermes and
Coralie are fibre-fed, optical high-resolution spectrographs, mounted on twin
telescopes. Hermes observes the range 377 -- 900\,nm at a spectral resolution of
$R = \lambda / \Delta \lambda = 85\,000$, whereas Coralie approximately covers
the range 385 -- 690\,nm, with a somewhat lower spectral resolution of
$R = 50\,000$. In terms of transmission, Hermes is by far the more efficient of
the two instruments.

The sample stars and the log of the observations is presented in
Table~\ref{obslog}.
%The instrument used is listed in column 2: ``H'' stands for Hermes, ``C'' for
%Coralie. The date of the observation is the mean date of the exposures used
%to construct a combined spectrum. The integration times of the individual
%exposures are listed in column 4 of that table.
We aimed to observe the stars close to their visual maximum brightness across
their pulsation cycle, to maximise the signal-to-noise ratio (SNR) of the
spectra.

The observed 2D spectra were reduced with the respective instrument pipeline
software. As a second measure to increase the SNR in the blue spectral range,
we combined the individual 1D spectra into one. This was essential for some of
the stars, as they have a very low flux in the blue range.
%The last column in Table~\ref{obslog} lists the estimated SNR of the combined
%spectrum around 430\,nm, which is close to the wavelength of the Tc lines.
Despite these efforts to increase the SNR, most stars have a low to moderate
SNR at 430\,nm. \object{W~Dra} is on average the faintest star in the sample
and could not be observed close to its visual brightness maximum. As a result,
the SNR of its spectrum in the $B$-band is essentially zero. Nevertheless, the
signal in the red part of the spectrum was high enough to search for other
signatures of s-process enrichment (see below). The SNR at 671\,nm, where the
Li resonance doublet lies, is about 25 in W~Dra's combined spectrum, which is
just good enough to determine the Li abundance. For the other stars, the SNR in
this spectral range is very high ($> 200$), so we used one of the individual
spectra instead of a combined spectrum to measure the Li abundance (see
Sect.~\ref{sectli}).

%-------------------------------------------------------------------------------
\begin{table}
\caption{Observation log.}
\label{obslog}
\begin{tabular}{lccrc}
\hline\hline
Name             & Instr. & Date         & Exp.\ time          & SNR        \\
                 &        & (dd/mm/yyyy) & (s)                 & (at 430\,nm) \\
\hline
\object{BH Cru}  & C      & 03/08/2009   & $4\times1800$       & ~12        \\
\object{RU Vul}  & H      & 04/07/2009   & $300 + 2\times600$  & ~30        \\
\object{T UMi}   & H      & 04/07/2009   & $4\times1800$       & ~20        \\
\hline
\object{R Aql}   & H      & 04/07/2009   & $900 + 2\times300$  & ~50        \\
\object{R Hya}   & H      & 13/02/2010   & $2\times400$        & ~60        \\
\object{W Dra}   & H      & 15/08/2009   & $3\times1800$       & ~$\sim$0   \\
\hline
\object{R Nor}   & C      & 17/08/2010   & $2\times1800$       & ~20        \\
\object{S Her}   & H      & 22/10/2009   & $630 + 2\times1200$ & ~30        \\
\object{S Ori}   & H      & 13/02/2010   & $3\times1800$       & ~15        \\
\object{T Cep}   & H      & 25/10/2009   & $2\times1200$       & 140        \\
\object{W Hya}   & H      & 13/02/2010   & $3\times150$        & ~50        \\
\hline
\object{T Her}   & H      & 04/07/2009   & $2\times800$        & ~70        \\
\hline
\end{tabular}
{\bf Note:} The meaning of the columns is as follows: (1) Name listed in
Simbad;  (2) Instrument used (H: Hermes, C: Coralie); (3) Mean date of the
observations; (4) Exposure times of the spectra in seconds; (5) Approximate SNR
at 430\,nm.
\end{table}
%-------------------------------------------------------------------------------

In Table~\ref{obslog}, the stars are grouped into the different period change
classes described in the introduction, and separated by horizontal lines: The
first group is that of the sudden change stars (SCh), the second one that of the
continuous change stars (CCh), and the third one the meandering Miras (MCh). The
star \object{T~Her} in the last row is an exception here. It is a Mira variable
with a constant period, hence it does not belong to any of the period change
classes. However, it is discussed among other stars in \citet{WZ81}, and it was
hitherto uncertain whether this star contains Tc. \citet{Lit87} classify it as
``Tc doubtful'', and we wish to clarify this situation. Furthermore,
\object{RU~Vul} and \object{W~Hya} are actually classified as SRVs in the
Simbad database. This classification is controversial, as we will further
elaborate on in Sect.~\ref{Discindiv}.
%A further exception is \object{RU~Vul}, which is actually a semi-regular
%variable (SRV), not a Mira. Its declining period was noted by \citet{ZB02}, and
%\citet{TWF08} found that its period declined from $\sim$155 to $\sim$108\,d
%between 1954 and 2007, while its amplitude has almost vanished by now. We
%include it in our sample because of its significant period change, reminiscent
%of the SCh Miras.
% Typo in the coordinates of RU Vul in TWF08!

\section{Analysis and Results}\label{AnalRes}

\subsection{Analysis of the spectra}

\subsubsection{Oxygen-rich versus carbon-rich}\label{secCO}

\begin{figure}
  \centering
  \includegraphics[width=\linewidth,bb=83 370 551 700]{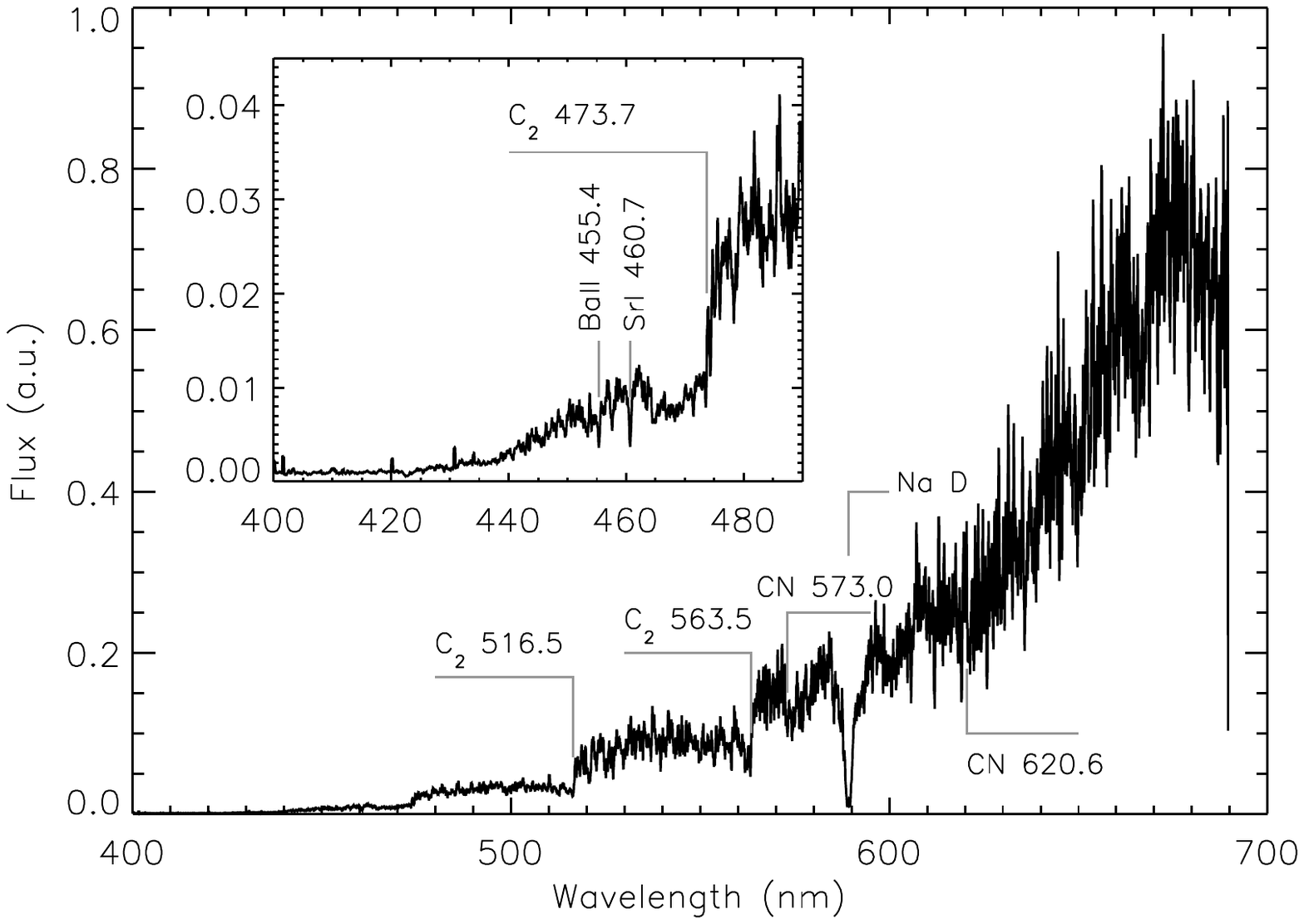}
  \caption{Coralie spectrum of \object{BH~Cru}, smoothed to a resolution of
    $\sim$0.3\,nm per wavelength point. The main features of the spectrum are
    identified. The inset shows a zoom on the spectrum blue-ward of 490\,nm,
    where the flux is very low.}
  \label{BHCruoverview}
\end{figure}

As a first step, we performed a global inspection of the spectra to decide
whether they are dominated by either features of oxygen-bearing molecules
(e.g.\ TiO, VO) or those of carbon-bearing molecules (e.g.\ C$_2$, CN). Stars
whose spectra are dominated by TiO and VO bands are oxygen-rich (i.e.\ they
have a C/O ratio $<1$ in their atmosphere), whereas stars whose spectra are
dominated by features of C$_2$ and CN are carbon-rich ($\rm{C}/\rm{O} > 1$). It
turns out that all but one star in our sample displays TiO bands, this one star
being \object{BH~Cru}. This star has prominent C$_2$ and (less prominent) CN
features in its spectrum, whereas ZrO bands are absent. Additional very
prominent features in BH~Cru include the Na\,D line, indicating a very low
temperature and/or C/O ratio, and the BaII\,4554 and SrI\,4607 lines,
indicating an s-process enrichment. Figure~\ref{BHCruoverview} shows the
Coralie spectrum of BH~Cru, smoothed to the resolution of classification
spectra to make the main features visible. The C$_2$ bands clearly dominate the
spectrum.

Among the other stars, the one with clearly the weakest TiO features is
\object{RU~Vul}, implying that it is the hottest star in our sample (see also
Sect.~\ref{sectli}). All other O-rich stars display pronounced bands of TiO.
The range of spectral types found in the compilation of \citet{Skiff} is
listed in column 3 of Table~\ref{tabres}. All stars exhibit hydrogen emission
lines from time to time.

%-------------------------------------------------------------------------------
%-------------------------------------------------------------------------------

\subsubsection{Technetium content}\label{tccont}

Before searching for Tc, it is necessary to measure the radial velocity of the
stars. To this end, we applied a cross-correlation technique, using a synthetic
model spectrum as a template. The spectral range of 430 -- 435\,nm was used to
complete this, except for W~Dra, for which only the red part of the spectrum
had enough signal and we then used the range 701 -- 710\,nm. The found
heliocentric radial velocities are summarised in column 4 of Table~\ref{tabres}.

\begin{figure}
  \centering
  \includegraphics[width=\linewidth,bb=83 370 536 699]{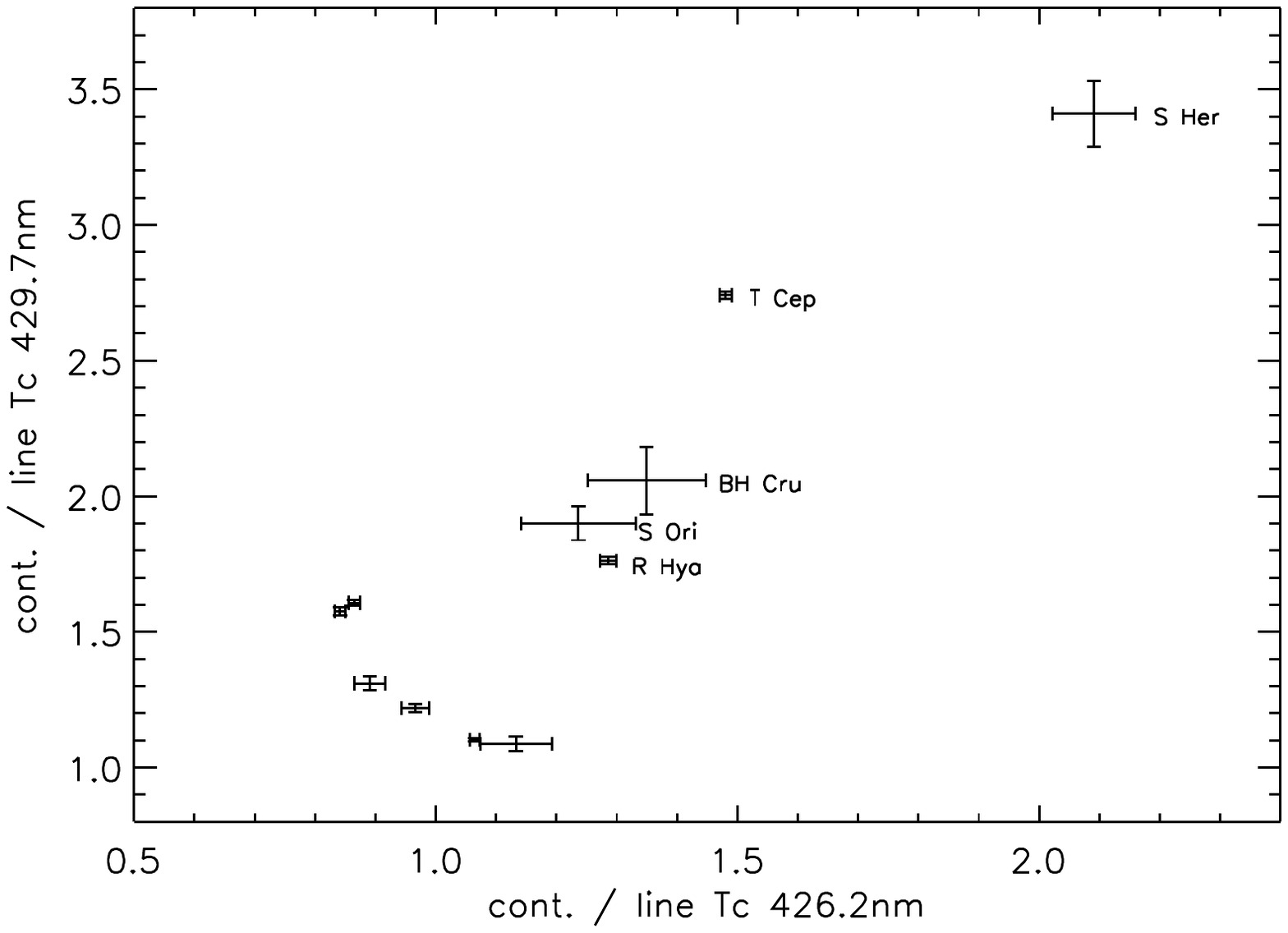}
  \caption{Continuum-to-line flux ratios of the Tc lines at 426.2 and 429.7\,nm
    of our sample stars. The data points of the Tc-rich stars, i.e.\ those stars
    with an enhanced ratio in both lines, are marked with their names.}
  \label{Tcfluxratio}
\end{figure}

With the radial velocity information, we determined whether or not Tc is present
in the sample stars using the flux-ratio method introduced by \citet{Utt07a},
and by visual inspection of the spectra around the strongest lines of neutral Tc
($\lambda = 398.497$, 403.163, 404.911, 409.567, 423.819, 426.227, and
429.706\,nm, in air). In this method, the mean flux contained within a range
of a few wavelength pixels centred on a quasi-continuum point is divided by the
mean flux in a small region around the Tc line. If Tc line absorption is
present, the resulting ratio will be considerable larger than one, but close to
one otherwise. The measured flux ratios for the two Tc lines at the longest
wavelengths, hence highest SNR, are shown in Fig.~\ref{Tcfluxratio}. For the Tc
426.2\,nm line, the range 426.120 -- 426.140\,nm was used for the continuum
point and 426.220 -- 426.240\,nm for the line, while for the Tc 429.7\,nm line
the ranges were 429.625 -- 429.655 and 429.690 -- 429.718\,nm, respectively.
The error bars of the data points were determined with a Monte Carlo simulation.
One-hundred trial spectra were created by adding Gaussian noise with the
magnitude of the inverse SNR to the observed, combined spectrum. The flux
ratios were measured from those trial spectra, and the standard deviation of
the distribution of the ratios was adopted as the error bar.

%-------------------------------------------------------------------------------
%-------------------------------------------------------------------------------

Five stars were found to have a higher continuum-to-line flux ratio, which was
confirmed by visual inspection of the spectra around the Tc lines. These stars
are \object{BH~Cru}, \object{R~Hya}, \object{S~Her}, \object{S~Ori}, and
\object{T~Cep}. The surprise in this list is R~Hya, which was previously
classified by \citet{UL10} as Tc-poor on the basis of a UVES spectrum of that
star. After careful inspection of all Tc lines in the observed range and a
comparison between the UVES and the Hermes spectrum of R~Hya, it became clear
that this star does indeed have Tc in its atmosphere, in contrast to the
findings of \citet{UL10}. The UVES observations on 6 July 2000 were performed
near maximum light of R~Hya, while the Hermes observations were done about
half-way between a maximum and a minimum. A cross-correlation of the two
spectra shows that there are velocity components in the atmosphere that differ
by up to 12\,km\,s$^{-1}$. The lines in the Hermes spectrum also appear deeper,
which is not simply a consequence of the somewhat lower resolution of the UVES
spectrum (50\,000, compared to 85\,000 for the Hermes spectrum). Many lines in
the UVES spectrum, including the Tc lines, seem to suffer from the often cited
``line weakening'' effect in Mira variables, which causes many spectral lines
to become much weaker at times, or to disappear completely \citep{MDK62}. This
seems to have happened with the near-maximum UVES spectrum of R~Hya, which we
have to re-classify as a Tc-rich star. This can also be seen as a warning that
intrinsically weak lines (the Tc lines of R~Hya are already at the weak end of
the Tc line strengths observed in the current sample) may go entirely
undetected if a Mira star is observed at a pulsation phase with severe line
weakening. There was no error in the analysis of \citet{UL10}: revisiting the
spectrum now, we would still classify it as Tc-poor. In view of this situation,
we scrutinized the present sample to assess whether we could have mistaken a
Tc-rich star for a Tc-poor one, but conclude that this is not the case.
%We are also sure that the same object has been observed with both instruments.

%Could we mistake a Tc-rich star for a Tc-poor star in the present sample
%because of line weakening? As already noted by \citet{Mer46,Mer52} in a series
%of spectra of \object{R~Hya}, the absorption lines of neutral metals are the
%strongest near minimum visual light, and the weakest near maximum visual light.
%Conversely, the hydrogen emission lines are the strongest near maximum visual
%light.
%%The atmospheric structure of Mira variables deviates most from
%%hydrostatic equilibrium around maximum visual light \citep{McS07}, as is also
%%certified by the appearance of the hydrogen Balmer lines in emission.
%Thus, we have to be cautious with the Tc-poor stars that have been observed
%around their maximum. This is the case for \object{T~UMi}, \object{W~Hya}, and
%\object{T~Her}. The hydrogen Balmer lines are not seen in emission in the
%spectra of T~UMi and W~Hya, thus the classification of these stars as Tc-poor
%should be unproblematic. However, T~Her shows prominent H emission lines, so
%this one could be questionable. \citet{Lit87} conclude that it is ``doubtful''
%that this star has Tc, which supports our classification. As T~Her has a
%constant pulsation period, hence does not belong to any of the period change
%groups, a wrong classification of this star would have the least effect on our
%conclusions. We may thus summarise that the main conclusions of the present
%study are not affected by the line weakening effect in Mira stars.

\begin{figure}
  \centering
  \includegraphics[width=\linewidth,bb=88 371 533 695]{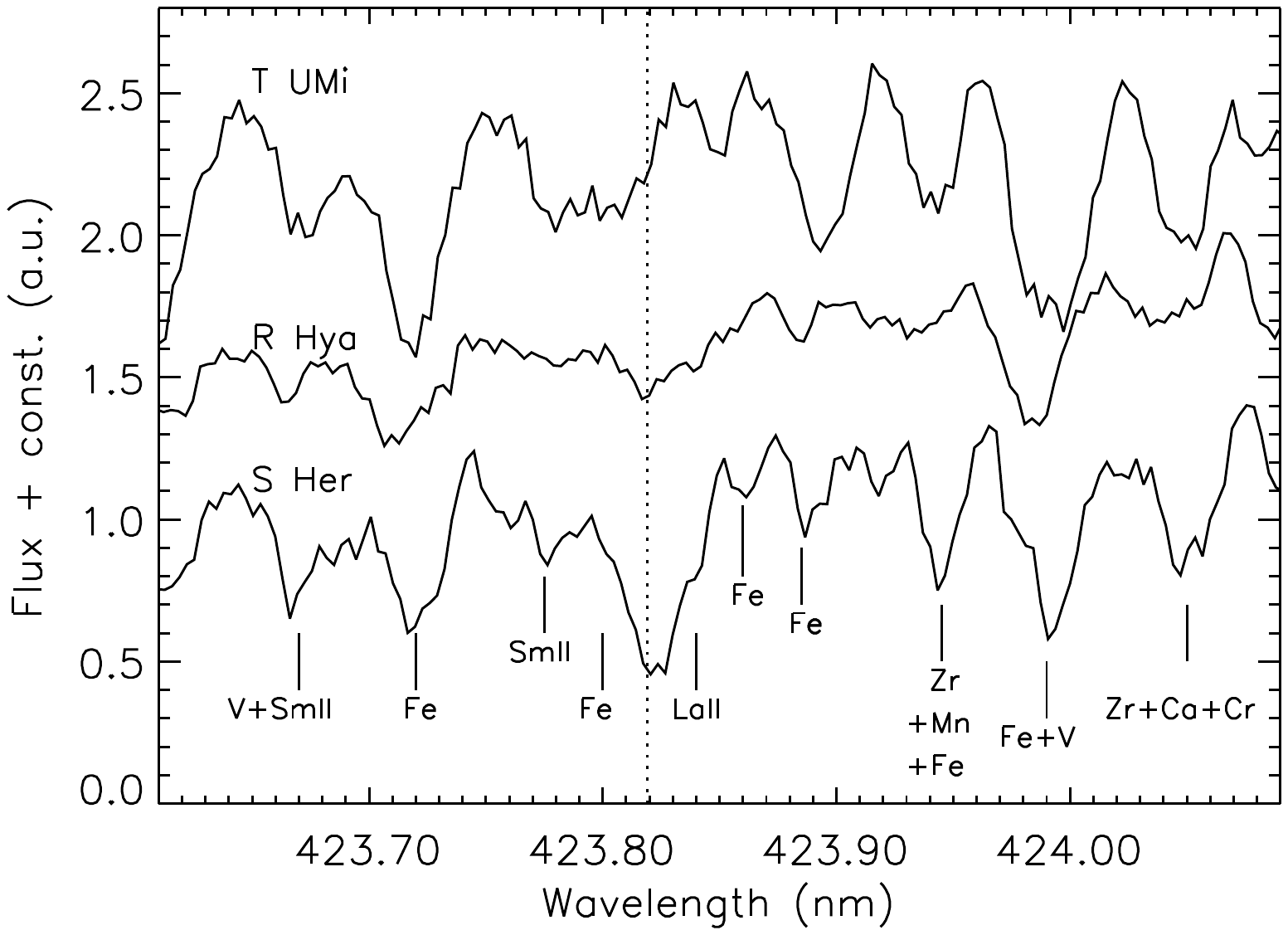}
  \caption{Spectra of three sample stars around the Tc line at 423.819\,nm. From
    top to bottom: \object{T~UMi}, \object{R~Hya}, and \object{S~Her}. The
    spectra of T~UMi and R~Hya have been shifted upwards by 1.2 and 0.6 to make
    the plot clearer. The wavelength of the Tc line is marked with a dotted
    vertical line.}
  \label{Tclines}
\end{figure}

In Fig.~\ref{Tclines}, the spectra of three sample stars around the Tc line at
423.819\,nm are shown. The three stars are chosen to delineate a sequence of
increasing Tc line strength from absent in \object{T~UMi}, weak in
\object{R~Hya}, and strong in \object{S~Her}. The lines in R~Hya are generally
weaker than in the other two stars, again illustrating the difficulty of
detecting the Tc lines in this star. A few of the surrounding lines are
identified, with identifications being taken from \citet{Davis}.

%-------------------------------------------------------------------------------
%-------------------------------------------------------------------------------

The results of our search for Tc are listed in column 5 of Table~\ref{tabres}.
Column 6 summarises what was previously reported in the literature about the Tc
content of the sample stars. This shows that the Tc content of most stars in
our sample was unclear, with there even being contradictory classifications. As
discussed above, this reflects that the search for the weak Tc lines in
crowded, low-SNR spectra of red giant stars is not a simple and
straight-forward task.

%-------------------------------------------------------------------------------
\begin{table*}
\caption{Results of our analysis.
%Details are to be found in Sect.~\ref{AnalRes}.
}
\label{tabres}
\begin{tabular}{lccrcccccccccc}
\hline\hline
Name             & Change   & Sp.\    &  $rv_{\rm{helio}}$   & Tc  & Tc lit.\      & $T_{\rm{eff}}$  & $\log \epsilon (\rm{Li})$ & $d \ln P / dt$         & $P$      & $P$   & $J-K$   & $M_{\rm{K}}$ & $M_{\rm{bol}}$ \\
                 & type     & type    & (km\,s$^{-1}$)     &     &               & (K)           &                           & ($10^{-3} \rm{yr}^{-1}$) & (d, T05) & (d)   &         &            &              \\
\hline%%%%%%%%%%%%%%%%%%%%%%%%%%%%%%%%%%%%%%%%%%%%%%%%%%%%%%%%%%%%%%%%%%%%%%%%%%%%%%%%%%%%%%%%%%%%%%%%%%%%%%%%%%%%%%%%%%%%%%%%%%%%%%%%%%%%%%%%%%%%%%%%%%%%%%%%%%%%%%%%%%%%%%%%%%%%%%%%%%%%%%%%%%%%%%%%%%%%%%%%%%
(1)              & (2)      & (3)     & (4)               & (5) & (6)           & (7)           &  (8)                      &  (9)                   & (10)     & (11)  & (12)    & (13)       & (14)         \\
\hline%%%%%%%%%%%%%%%%%%%%%%%%%%%%%%%%%%%%%%%%%%%%%%%%%%%%%%%%%%%%%%%%%%%%%%%%%%%%%%%%%%%%%%%%%%%%%%%%%%%%%%%%%%%%%%%%%%%%%%%%%%%%%%%%%%%%%%%%%%%%%%%%%%%%%%%%%%%%%%%%%%%%%%%%%%%%%%%%%%%%%%%%%%%%%%%%%%%%%%%%%%
\hline%%%%%%%%%%%%%%%%%%%%%%%%%%%%%%%%%%%%%%%%%%%%%%%%%%%%%%%%%%%%%%%%%%%%%%%%%%%%%%%%%%%%%%%%%%%%%%%%%%%%%%%%%%%%%%%%%%%%%%%%%%%%%%%%%%%%%%%%%%%%%%%%%%%%%%%%%%%%%%%%%%%%%%%%%%%%%%%%%%%%%%%%%%%%%%%%%%%%%%%%%%
\object{BH Cru}  & SCh      & C       & ~$-3.7$           & yes & \dots         & 3000          & --0.7                     & $+3.70$                & ~~530    & 524.4 & 1.644   & $-8\fm69$  & $-5\fm59$    \\
\object{RU Vul}  & SCh      & M2-4    & $-67.9$           & no  & \dots         & 3700          & $\lesssim$--1.5           & $-7.11$                & ~~112    & 108.8 & 1.223   & $-6\fm88$  & $-3\fm89$    \\
\object{T UMi}   & SCh      & M4-6    & ~$+8.7$           & no  & \dots         & 3300          & $\lesssim$0.0             & $-8.47$                & ~~240    & 229.1 & 1.553   & $-7\fm44$  & $-4\fm36$    \\
\hline%%%%%%%%%%%%%%%%%%%%%%%%%%%%%%%%%%%%%%%%%%%%%%%%%%%%%%%%%%%%%%%%%%%%%%%%%%%%%%%%%%%%%%%%%%%%%%%%%%%%%%%%%%%%%%%%%%%%%%%%%%%%%%%%%%%%%%%%%%%%%%%%%%%%%%%%%%%%%%%%%%%%%%%%%%%%%%%%%%%%%%%%%%%%%%%%%%%%%%%%%%
\object{R Aql}   & CCh      & M5-9    & $+59.3$           & no  & prob$^{\rm{1}}$ & 3000          & $\lesssim$0.0             &   $-1.56$              & ~~270    & 270.6 & 1.361   & $-7\fm68$  & $-4\fm81$    \\
                 &          &         &                   &     & dbfl$^{\rm{2}}$ &               &                           &                        &          &       &         &            &              \\
                 &          &         &                   &     & no$^{\rm{3}}$   &               &                           &                        &          &       &         &            &              \\
\object{R Hya}   & CCh      & M6-9    & $-15.9$           & yes & yes$^{\rm{4}}$  & 3100          & $\lesssim$0.0             &   $-0.71$              & ~~385    & 376.6 & 1.261   & $-8\fm16$  & $-5\fm17$    \\
                 &          &         &                   &     & prob$^{\rm{2}}$ &               &                           &                        &          &       &         &            &              \\
                 &          &         &                   &     & no$^{\rm{5}}$   &               &                           &                        &          &       &         &            &              \\
\object{W Dra}   & CCh      & M3-4    & ~$-3.6$           & ?   & \dots         & 2950          & $\lesssim$0.0             &   $+1.03$              & ~~282    & 289.6 & 1.340   & $-7\fm78$  & $-4\fm74$    \\
\hline%%%%%%%%%%%%%%%%%%%%%%%%%%%%%%%%%%%%%%%%%%%%%%%%%%%%%%%%%%%%%%%%%%%%%%%%%%%%%%%%%%%%%%%%%%%%%%%%%%%%%%%%%%%%%%%%%%%%%%%%%%%%%%%%%%%%%%%%%%%%%%%%%%%%%%%%%%%%%%%%%%%%%%%%%%%%%%%%%%%%%%%%%%%%%%%%%%%%%%%%%%
\object{R Nor}   & MCh      & M3-7    & $-28.5$           & no  & \dots         & 3200          & $+4.6$                    & $+0.29$                & ~~510    & 496.2 & 1.310   & $-8\fm55$  & $-5\fm50$    \\
\object{S Her}   & MCh      & M4-7.5S & $-17.5$           & yes & yes$^{\rm{2}}$  & 3300          & $\lesssim$0.0             & $+0.06$                & 307.5    & 304.1 & 1.166   & $-7\fm85$  & $-4\fm89$    \\
\object{S Ori}   & MCh      & M5-10   & $+34.5$           & yes & dbfl$^{\rm{2}}$ & 2950          & $+0.8$                    & $+0.35$                & ~~430    & 433.4 & 1.352   & $-8\fm39$  & $-5\fm49$    \\
\object{T Cep}   & MCh      & M5-9    & $-15.7$           & yes & prob$^{\rm{1}}$ & 3000          & $+0.5$                    & $+0.12$                & 390.9    & 386.6 & 1.328   & $-8\fm19$  & $-5\fm16$    \\
                 &          &         &                   &     & yes$^{\rm{2}}$  &               &                           &                        &          &       &         &            &              \\
\object{W Hya}   & MCh      & M7-9    & $+40.5$           & no  & prob$^{\rm{1}}$ & 3300          & $+0.5$                    & \dots                  & ~~361    & 388.6 & 1.459   & $-8\fm20$  & $-5\fm25$    \\
                 &          &         &                   &     & no$^{\rm{6}}$   &               &                           &                        &          &       &         &            &              \\
\hline%%%%%%%%%%%%%%%%%%%%%%%%%%%%%%%%%%%%%%%%%%%%%%%%%%%%%%%%%%%%%%%%%%%%%%%%%%%%%%%%%%%%%%%%%%%%%%%%%%%%%%%%%%%%%%%%%%%%%%%%%%%%%%%%%%%%%%%%%%%%%%%%%%%%%%%%%%%%%%%%%%%%%%%%%%%%%%%%%%%%%%%%%%%%%%%%%%%%%%%%%%
\object{T Her}   & const.\ & M2-8    & $-91.9$            & no  & dbfl$^{\rm{2}}$ & 3400          & $+2.0$                    & $+0.04$                & 164.9    & 163.9 & 1.222   & $-6\fm96$  & $-4\fm20$    \\
\hline%%%%%%%%%%%%%%%%%%%%%%%%%%%%%%%%%%%%%%%%%%%%%%%%%%%%%%%%%%%%%%%%%%%%%%%%%%%%%%%%%%%%%%%%%%%%%%%%%%%%%%%%%%%%%%%%%%%%%%%%%%%%%%%%%%%%%%%%%%%%%%%%%%%%%%%%%%%%%%%%%%%%%%%%%%%%%%%%%%%%%%%%%%%%%%%%%%%%%%%%%%
\end{tabular}
{\bf Note:} The meaning of the columns is as follows: (1) Name listed in Simbad;
(2) Period change type; (3) Range of spectral type; (4) Heliocentric radial
velocity; (5) Presence of Tc, this paper; (6) Presence of Tc, previous
studies; (7) Effective temperature; (8) Abundance of Li; (9) Rate of period
change; (10) Period as given in \citet{TMW05}; (11) Period found in this
study; (12) Mean $J - K$ colour; (13) Absolute $K$ magnitude; (14) Bolometric
magnitude. References of previous studies on Tc content: 1: \citet{LML79}, 2:
\citet{Lit87}, 3: \citet{Van91}, 4: \citet{Mer52}, 5: \citet{UL10}, 6:
\citet{LH99}. Abbreviations mean: ``prob'' = probably, ``dbfl'' = doubtful.
\end{table*}
%-------------------------------------------------------------------------------

\subsubsection{Other s-process indicators}\label{sindic}

The spectrum of \object{W~Dra} has basically no signal at the wavelength of the
Tc lines, hence we searched its spectrum for other indicators of s-process
enrichment. The bands of the ZrO molecule are such an indicator, because Zr is
also an s-process element. The formation of ZrO is also facilitated by an
enhanced C/O ratio, because at $\rm{C}/\rm{O} \lessapprox 1$ the molecular
equilibrium is shifted to ZrO and LaO, at the expense of TiO and VO. However,
in contrast to Tc, Zr is a stable element, hence is not an indicator of a
{\em recent} s-process and 3DUP. An enhanced Zr abundance can also be caused
e.g.\ by the mass transfer of s-process enriched material from a binary
companion, which would now be a white dwarf. The mass transfer could have
happened billions of years ago.While Tc would have decayed by now, Zr would
still be present in the atmosphere of the now evolved secondary star. This
binary mass transfer scenario is the probable mechanism leading to Tc-poor
S-type stars \citep{VEJ99}. The absence of Zr enrichment (and hence of ZrO
bands) would however indicate that neither intrinsic nor extrinsic s-process
enrichment has taken place in a given star.

\begin{figure}
  \centering
  \includegraphics[width=\linewidth,bb=83 368 546 700]{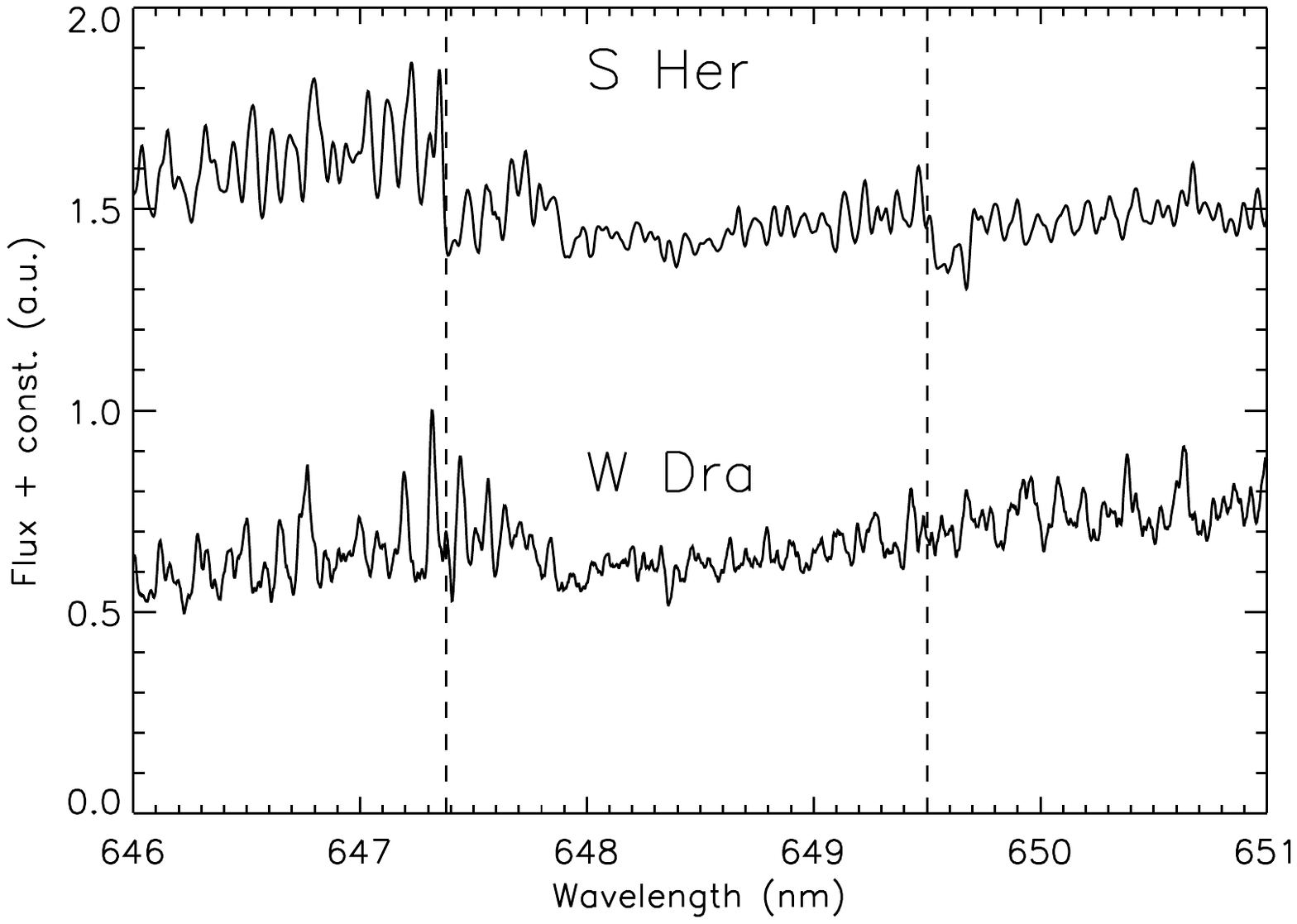}
  \caption{Spectra of \object{S~Her} (top) and \object{W~Dra} (bottom) around
    the ZrO bandheads at 647.4 and 649.5\,nm, smoothed by a nine-point running
    box-car. The theoretical wavelengths of the bandheads are marked by dashed
    lines. The feature at 647.9\,nm is a TiO bandhead.}
  \label{ZrOexample}
\end{figure}

Particularly strong and weakly blended bandheads of ZrO are to be found in cool
star spectra at wavelengths of 647.4 and 649.5\,nm. The combined spectrum of
\object{W~Dra} has enough flux at this wavelength to detect the ZrO bands if
they were present, but apparently they are not (Fig.~\ref{ZrOexample}). In
addition, the spectra of all other sample stars were inspected for the ZrO
bands, but only \object{S~Her} definitely shows them\footnote{As discussed in
Sect.~\ref{secCO}, \object{BH~Cru} is C-rich, which leaves no trace of
O-bearing molecules such as ZrO.}. As seen from Fig.~\ref{Tcfluxratio}, S~Her
is also the star with the strongest Tc lines in our sample, so it is clearly of
MS spectral type. Since ZrO is not detectable in three of the Tc- and O-rich
sample stars, the absence of ZrO bands does {\em not} imply the absence of Tc.
With ZrO being absent in W~Dra, we cannot draw a firm conclusion about the Tc
content of this star.
%Note that a too high atmospheric temperature for ZrO to form or a C/O ratio
%very close to unity would let the ZrO bands disappear, too. However, also the
%bands of TiO would disappear under either of these conditions, which is
%clearly not the case.
We can only conclude that W~Dra is an O-rich, M-type star, without strong
s-process (and C/O) enhancement.
% As an illustration of the above
%said, Fig.~\ref{ZrOexample} shows the spectra of W~Dra and S~Her around the ZrO
%band heads at 647.4 and 649.5\,nm.

%-------------------------------------------------------------------------------
%-------------------------------------------------------------------------------

\subsubsection{Lithium abundance}\label{sectli}

In contrast to Tc, we can determine for Li not only its presence but also its
abundance, because state-of-the-art model spectra of cool giants are able to
reproduce observed spectra much more closely in the region of the Li line than
around the Tc lines. The most important stellar parameter that determines the
strength of the Li\,I resonance doublet at 670.8\,nm is the effective
temperature. For the O-rich stars in our sample, the temperature can be
determined from the strength of the TiO bands.
%Because the TiO molecule dissociates at high temperatures, the bands are very
%sensitive to the temperature.
Our method for temperature determination involves a grid of COMARCS model
atmospheres and spectra \citep{Ari09}, and closely follows the approach of
\citet{GH07}. %{\bf For details of the method we refer to \citet{UL10}.}
Details of the method can be found in \citet{UL10}, and here we only outline the
main steps of the procedure.

A grid of synthetic spectra based on model atmospheres with temperatures
between 2600 and 3600\,K was calculated. The step width in temperature was
100\,K, and two models were added at 2950 and 3150\,K. A number of other
parameters, which have a much smaller impact on the spectrum than the
temperature, have been fixed beforehand to realistic values. The adopted
parameter values are: $M = 1 M_{\odot}$, $\log g [\rm{cm\,s}^{-2}] = 0.0$,
$Z = Z_{\odot} = 0.0156$, C/O=0.48 (solar\footnote{A C/O ratio fixed to the
solar value can be problematic for MS- or S-stars. However, we have only one
MS-star in our sample (\object{S~Her}, which shows ZrO bands), and it turns out
that its Li line is undetectable, independent of the assumed C/O ratio.}), and
micro-turbulence $\xi = 2.5$\,km\,s$^{-1}$. The spectra were calculated for
regions that contain band heads of the TiO molecule, namely 668.4 -- 674.0
(which also includes the Li line itself) and 701 -- 710\,nm. Before fitting the
synthetic spectra to the observed ones, they were smoothed with a Gaussian
kernel to the resolution of the respective instrument. An additional
macro-turbulence had to be adopted, with the maximum value of 9\,km\,s$^{-1}$
for \object{R~Nor}. The synthetic spectra were then fit to the observed spectra
using a $\chi^2$ minimisation method. The temperature of the model spectrum
that most closely reproduced a given observed spectrum was adopted as the
effective temperature of that star. These temperatures are listed in column 7
of Table~\ref{tabres}.

Three exceptions to this general procedure have to be noted. The Coralie
spectrum of \object{R~Nor} does not reach beyond 690\,nm, thus the temperature
was determined only from the region 668.4 -- 674.0\,nm. \object{RU~Vul} has too
weak TiO bands to reliably determine the temperature from them. From its mean
spectral type (M3) and the calibration in \citet{Flu94}, a temperature of
3666\,K can be inferred.
%We also applied the $T_{\rm{eff}}$ vs.\ $V-K$ relation of \citet{vanBelle99} to
%the data of RU~Vul. Using the 2MASS $K$ magnitude {\bf (4\fm704)} and the mean
%$V$ magnitude of the AAVSO light curve {\bf (8\fm905)}, we find $V-K = 4.201$,
%which corresponds to 3808\,K in the calibration of \citet{vanBelle99},
%somewhat higher than the temperature derived from the spectral type.
% With dereddening for d=2684 pc and Av=0.7mag/kpc, I get (V-K)0=4656K, much
% hotter than is plausible for a spectral type M3!
We adopted 3700\,K for the spectral synthesis. At this temperature, the TiO
bands predicted by the synthetic spectra (which are based on solar metallicity)
are much too strong. However, the temperature may also not be much higher than
this because the TiO bands, the classification criterion of M stars, are
clearly present in RU~Vul, albeit weak.
% Fluks et al.: M0 (MK system) star has 3895 K.
The observed spectrum in the two spectral pieces can only be fit reasonably
well if we assume a metallicity 1.5\,dex below the solar value. A comparison
with the hotter, metal-poor giant Arcturus ($T_{\rm{eff}} = 4290$\,K,
$[\rm{M}/\rm{H}] = -0.5$) also shows that RU~Vul has much weaker metal lines.
Although we did not make a full metallicity determination, we suggest that
RU~Vul is a quite metal-poor star. Except for RU~Vul, the temperature range of
our sample stars is fairly narrow. The C-star \object{BH~Cru} does not exhibit
the temperature-sensitive TiO bands, so we adopted the temperature of 3000\,K
reported by \citet{Zij04}. For the spectral synthesis, we used model no.\ 443
from \citet{Ari09}, which has $T_{\rm{eff}}=3000$\,K, a C/O ratio of 1.05, and
otherwise identical parameters to our grid models. A C/O ratio higher than 1.05
does not significantly change the spectral appearance in the vicinity of the Li
line, hence has very little influence on the measured Li abundance.

In the next step, spectra with varying amounts of Li were synthesised based on
the best-fit model atmosphere. The hyperfine structure of the Li line was
taken into account in the spectral synthesis, but not the isotopic shift (i.e.\
only the contribution by $^7$Li was included). This effect can be neglected
here, because the $^6$Li abundance can be expected to be very small, and other
effects such as the (velocity) structure of the atmosphere, blending with TiO
lines, etc., play a much more important role. The Li abundance that yielded a
minimum in the residuals of observed -- calculated flux was adopted as the
measured Li abundance and is reported in column 8 of Table~\ref{tabres}. The
uncertainty in the Li abundance is however large, up to 0.6\,dex
\citep{GH07,UL10}. The largest part of this uncertainty is based on the
uncertainty in the temperature (0.4\,dex), the other stellar parameters and
flaws in the (TiO) line lists contribute to the rest.

Because of the large opacity of the TiO molecule relative to that of C$_2$ and
CN at the wavelength of the Li doublet (a factor of $\sim$10 difference at
$T_{\rm{eff}} = 3000$\,K!), the Li detection threshold for O-rich stars is much
higher than for C-rich stars. Hence, the abundance of
$\log \epsilon (\rm{Li}) = -0.7$ measured in \object{BH~Cru} would not be
detectable in M-type stars of similar temperature. \citet{KiWa90} measure a Li
abundance in BH~Cru of $-1.5$\,dex relative to the Sun, or $-0.4$ on the
$\log \epsilon$ scale, in fairly good agreement with our result. The good
agreement is remarkable because BH~Cru has changed its spectral type
considerably over the past few decades, and our spectrum was taken almost 26
years after the one used by \citet{KiWa90}. Furthermore, the synthetic spectrum
based on the model chosen for BH~Cru (C/O~=~1.05) closely reproduces the
molecular lines (C$_2$, CN) in the observed spectrum in the vicinity of the Li
line, which is additional support for a C/O ratio $>1$ in that star.

Irrespective of the metallicity of \object{RU~Vul}, the Li line in this star is
undetectable. Because of its higher temperature and less crowded spectrum, an
upper limit of $\log \epsilon (\rm{Li}) \lesssim -1.5$ appears reasonable.

\begin{figure}
  \centering
  \includegraphics[width=\linewidth,bb=77 368 541 703]{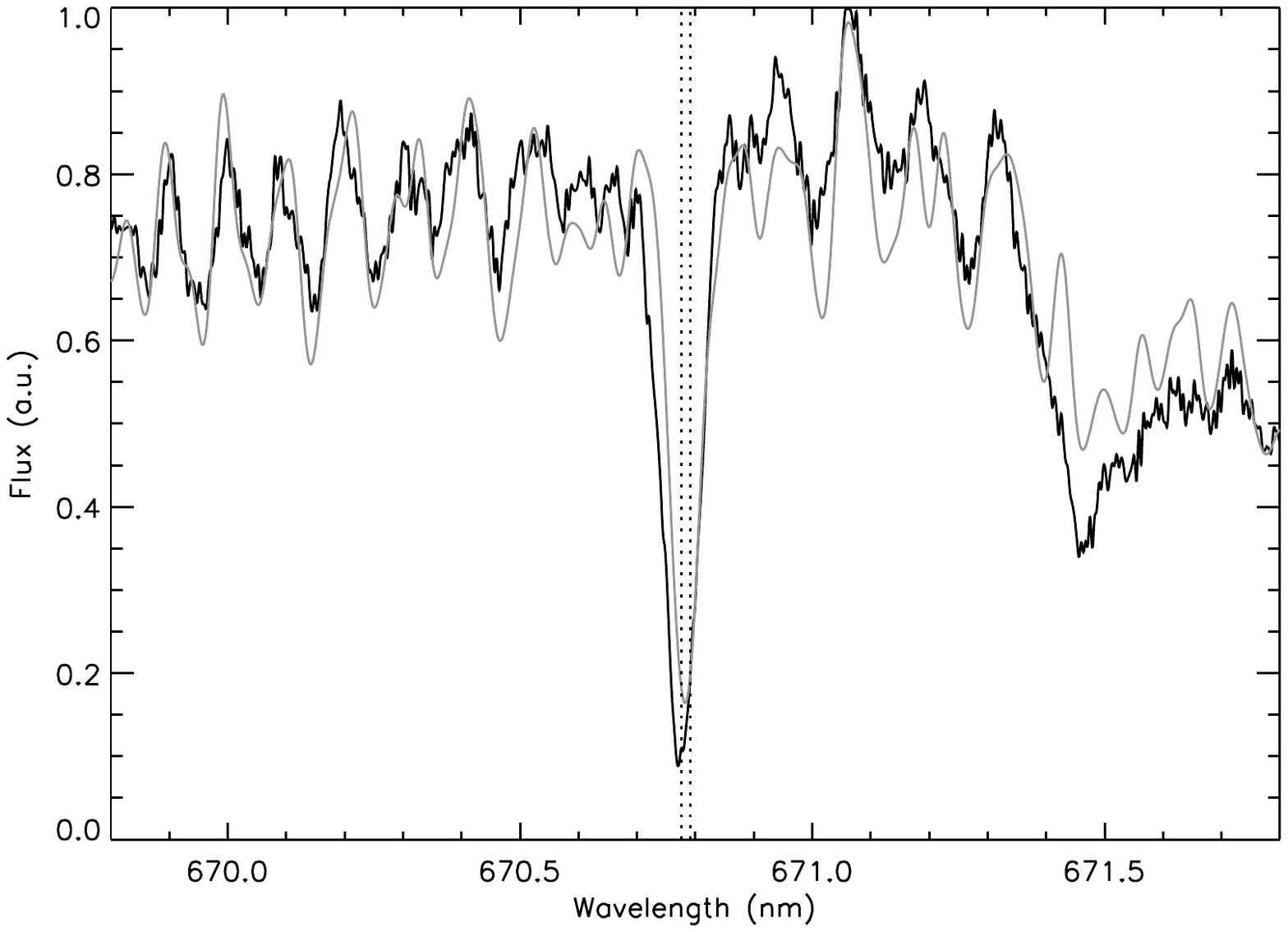}
  \caption{Observed spectrum of \object{R~Nor} around the Li line doublet (black
    graph), together with a synthetic spectrum (grey graph). The synthetic
    spectrum is based on a COMARCS model atmosphere with $T_{\rm{eff}} = 3200$\,K,
    $\log g = 0.0$, solar metallicity, and solar C/O ratio. A Li abundance of
    $\log\epsilon(\rm{Li}) = +4.6$ was assumed in the spectral synthesis. The
    dotted vertical lines indicate the laboratory wavelengths of the Li line
    doublet. All other features in the plotted range are due to TiO.}
  \label{rnorliline}
\end{figure}

The only period change class with noteworthy amounts of Li are the meandering
Miras. In this group, only \object{S~Her} does not appear to have Li, whereas
\object{R~Nor} has a huge amount of it. The measured abundance is
$\log \epsilon (\rm{Li}) = +4.6$, which is higher than the solar photospheric
abundance by a factor of $\sim$3200! The observed spectrum of R~Nor together
with a synthetic spectrum to fit the Li line is shown in Fig.~\ref{rnorliline}.
With the adopted abundance, only the red wing of the line is reproduced well,
whereas the blue wing would require an even higher abundance. However, some of
the absorption in the blue wing probably stems from circumstellar absorption
in the expanding stellar wind. This is supported by the profile of the NaI~D
doublet, which also exhibits a strong blue-shifted absorption component. We did
not attempt to fit the blue-shifted component in the Li line, because our model
atmospheres do not include circumstellar shells. The very high Li abundance in
R~Nor suggests that it is undergoing HBB, and that it is a fairly massive star.
R~Nor is in some respects similar to the very Li-rich S-star \object{V441~Cyg},
but is not s-process enriched \citep{UL10}.
%The Li line in R~Nor
%is extremely deep and broad, the flux in the line core goes down to about 9\%
%of the local quasi-continuum. The line is about as strong as in the star
%\object{V441~Cyg} \citep{UL10}. \citet{UL10} argue that V441~Cyg could owe its
%large amount of Li to the operation of HBB, which would qualify it as massive
%AGB star. Due to uncertainties in the pulsation mode (and hence luminosity) of
%V441~Cyg, this conclusion was not firm. Because of the high Li abundance and
%high luminosity of R~Nor, we propose that also this star is a massive AGB star
%that undergoes HBB, producing Li in large amounts by the Cameron-Fowler
%mechanism. This measured Li abundance is at the upper limit of the predicted
%range for AGB stars undergoing HBB.

%-------------------------------------------------------------------------------
%-------------------------------------------------------------------------------

%There is, however, a difference between \object{R~Nor} and \object{V441~Cyg}
%that needs to be noted. Contrary to V441~Cyg, R~Nor does not show any signs of
%s-process {\bf or carbon} enrichment, neither Tc nor ZrO. The lack of Tc in
%R~Nor can most easily be explained by the lack of 3DUP events. However, if it
%is a massive AGB star, it is also possible that the envelope contains so much
%mass that the dredged-up material becomes so diluted that Tc and ZrO remain
%below the detection threshold. R~Nor is in this respect very similar to the
%Li-rich, s-poor Galactic AGB stars identified by \citet{GH07}, and the argument
%put forward by these authors for the lack of ZrO bands also invokes a massive
%envelope.
%
%The lack of Tc and ZrO has one more consequence on the measured Li abundance in
%\object{R~Nor}. As noted by \citet{Ple03}, atomic lines become considerably
%stronger in giant spectra as C/O approaches unity, because of the disappearing
%molecular opacity. This was a major uncertainty in the Li abundance measurement
%of \object{V441~Cyg}, because no information on its C/O ratio was available. In
%R~Nor, however, we can safely assume that C/O has not been raised considerably
%from its primordial (about solar) value by the dredge-up of C, {\bf because of
%the absence of ZrO and LaO bands (which are formed at C/O ratios close to
%unity, at the expense of TiO)}. Hence, the strong Li line is very likely caused
%by a high Li abundance alone, and is not the result of a C/O ratio close to
%unity.

Interestingly, the Tc-poor, constant-period Mira \object{T~Her} also contains
a considerable amount of Li. It is thus similar to the Galactic bulge AGB star
\object{Plaut 3-45} discussed in \citet{Utt07b}. This star could be another
example of a low-mass AGB star experiencing extra-mixing, thereby producing Li.

\subsection{Present-day periods}\label{presdper}

Most of our sample stars have experienced a considerable change in pulsation
period in the past few decades and centuries. To check whether or not this
period change has continued in recent years, we analysed the AAVSO visual light
curves to determine the present-day periods. We downloaded from the AAVSO web
page\footnote{http://www.aavso.org} the visual and $V$-band observations between
JD 2\,452\,000 and JD 2\,455\,500, i.e.\ between 31 March 2001 and 31 October
2010, and completed a Fourier analysis on the unprocessed data with the program
period04 \citep{LB05}. The detected periods (in days) are summarised in column
11 of Table~\ref{tabres}. The peaks in the Fourier spectrum are fairly sharp,
the Monte Carlo simulation performed with period04 yielded period uncertainties
of or less than 1 day. Only the Fourier spectrum of \object{RU~Vul} has no
clearly highest peak. The one at 108.8\,d is only barely higher than other
peaks, and is broad and asymmetric. We thus conclude that no clear periodic
pulsation is present any more in this star.

\citet{TMW05}, who use AAVSO data up to about 11 September 2003, do not list
present-day periods, but instead list average periods over the entire AAVSO time
series available for a particular star (which can be between a few decades and
a century long). Nevertheless, they do provide diagrams of the period evolution
for the stars with the largest period change, which are useful to check whether
the period change has continued. The last period value plotted in these
diagrams is the one listed in column 10 of Table~\ref{tabres}.
%These periods are listed in column 10 of Table~\ref{tabres}.
The meandering Miras \object{S~Her} and \object{T~Cep} and the ``constant''
Mira \object{T~Her} were not found to have a large period change (on average
over the total time span of AAVSO data) by \citet{TMW05}, thus no such plots
are available for them, and we instead list the {\em average} period given in
\citet{TMW05}. \object{W~Hya} was not included in the survey of \citet{TMW05},
probably because it is classified as SRV in Simbad. This classification
is controversial, and we discuss it further in Sect.~\ref{Discindiv}.
%It is, however, not entirely clear whether this star is an SRV or a
%Mira. This question was discussed in \citet{Now10} in more detail. Indeed, from
%its light variation and location in the P-L diagram W~Hya would be classified
%as a Mira. From the velocity variations in its atmosphere, on the other hand,
%it has clearly to be assigned to the group of SRVs. \citet{Leb05} concluded
%that a high mass might be the reason for the remarkable velocity behaviour.
%This must be a mis-classification, because its amplitude in the $V$ band is
%$\Delta V \approx 4\fm0$ since the beginning of AAVSO observations (around
%1946), well above the Mira definition limit of $\Delta V = 2\fm5$.
In any case, we decided to list in Table~\ref{tabres} the period of W~Hya as
given in the General Catalogue of Variable Stars \citep{Sam09}. In this table
(column 9), we also indicate the rate of period change, $d \ln P / dt$, defined
as the slope of the period divided by the average period \citep{TMW05}. For
\object{RU~Vul}, this rate is taken from \citet{TWF08}, and for W~Hya the rate
is missing.

\begin{figure}
  \centering
  \includegraphics[width=\linewidth,bb=82 370 571 604]{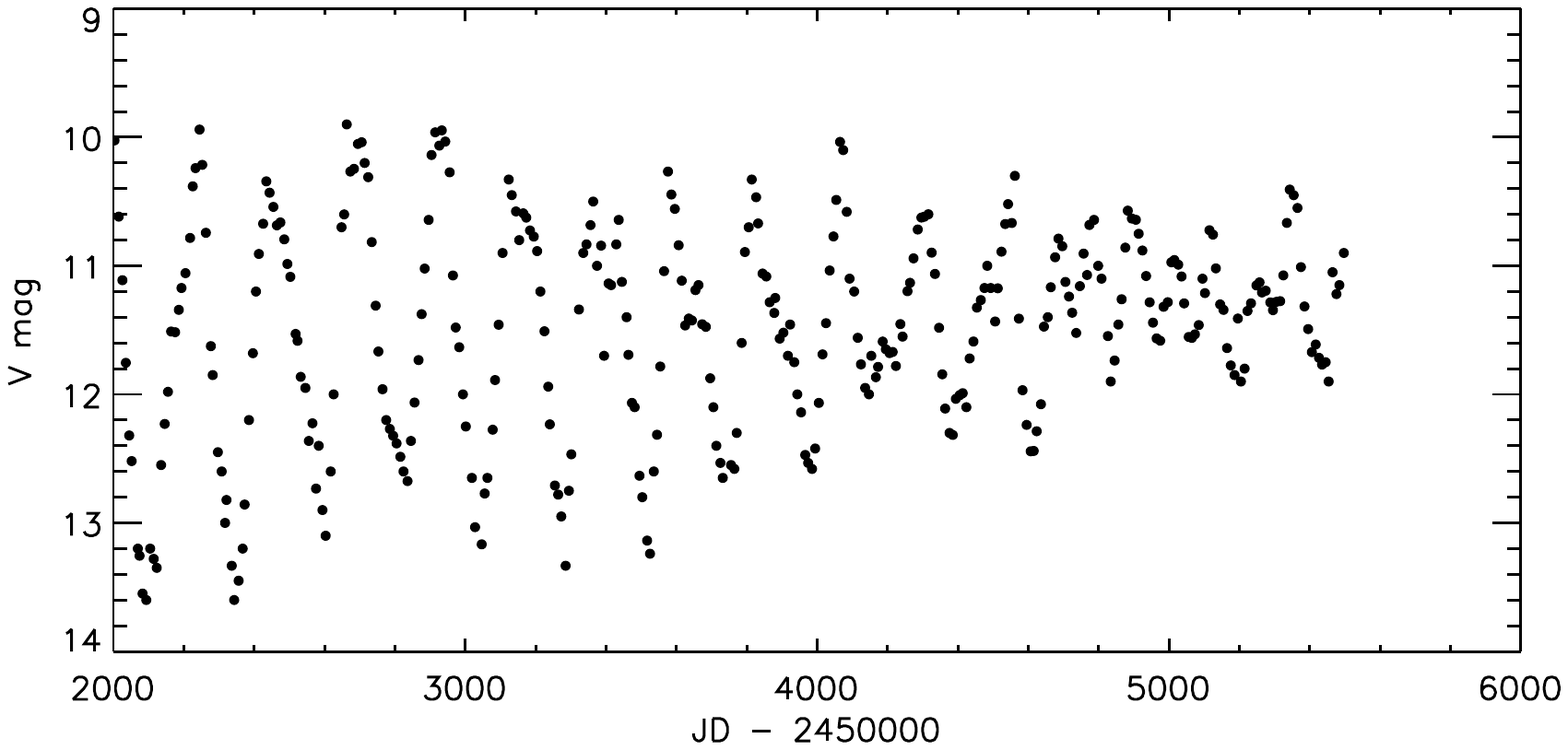}
  \caption{AAVSO light curve of \object{T~UMi} since JD 2\,452\,000. The data
    points are 10-day means. The previously dominant period of $\approx 230$\,d
    has disappeared in the most recent data.}
  \label{tumilc}
\end{figure}

From these data, we can make the following observations. Among the SCh change
type stars, \object{BH~Cru} seems to have stopped or maybe even reversed its
period increase. A deceleration of the period increase was proposed by
\citet{TMW05}. \object{RU~Vul} has continued its period decrease
\citep[also found by][]{TWF08}, and the pulsation is very weak now.
It is not detectable any more by visual observations in the most recent data.
The most dramatic evolution in recent years has however been seen in
\object{T~UMi}. In the time span investigated here, its dominant period
decreased to 229\,d and did not stop there. In the most recent data we see that
the star might even have switched its pulsation mode from Mira-like to
semi-regular variability around JD 2\,454\,600 (May 2008), as the dominant
period made a jump to 113.6\,d! A period ratio close to 2 can be expected for a
switch from fundamental mode (Mira-like) to first overtone (SRV) pulsation
\citep{Wood91}. At the same time, the full amplitude decreased to about
$1\fm5$. According to the classical (but arbitrary) definition of
$\Delta V > 2\fm5$, this small amplitude does not qualify T~UMi as a Mira any
more, but as SRV. The light curve of T~UMi in the time span investigated here
is displayed in Fig.~\ref{tumilc}. Continued photometric monitoring will be
very important to follow the exciting evolution of this star.

%-------------------------------------------------------------------------------
%-------------------------------------------------------------------------------

Among the CCh Miras, \object{R~Hya} has continued to decrease its period, and
\object{W~Dra} has continued to increase it, while \object{R~Aql}'s period
remained practically constant. This is somewhat in contradiction to the
luminosity evolution predicted for these stars by \citet{WZ81}, as the period
of W~Dra is expected to stop increasing at some point, whereas R~Hya and R~Aql
are expected to continue their decrease, albeit at a somewhat slower rate. The
development on these short timescales may however not be regarded as a stringent
constraint of the TP model employed by \citet{WZ81}. Among the MCh Miras,
\object{R~Nor} and \object{W~Hya} are the stars that display the strongest
changes. The former decreased its period in the recent years, while the latter
increased it considerably.
%\object{S~Her}, \object{S~Ori}, and \object{T~Cep} had fairly constant
%periods in the recent years, as was also expected for the ``constant'' Mira
%\object{T~Her}.

\subsection{Luminosities}\label{luminos}

To compare the evolutionary states in the HRD of the different period change
classes, as well as of the Tc-rich and the Tc-poor stars, we determined the
absolute bolometric magnitudes (luminosities) of our sample stars. In principle,
this could be done very easily using a period-luminosity relation for Mira
stars, e.g.\ the one from \citet[][their Fig.~A.1]{GB08}. However, our sample
stars do not necessarily closely follow this relation, because of the change in
the pulsation period. For example, \citet{Zij04} argue that the near-IR data of
\object{BH~Cru} is consistent with evolution at constant luminosity, and that
any increase in luminosity did not exceed 10 percent.
%However, Whitelock et al. (2006) find a brightening in $JHKL$ between
%1984 -- 1989 to 1997 -- 2004!
If this were true, then the star would now be fainter than other stars of a
similar pulsation period that do follow the Mira period-luminosity relation.
The only way to determine luminosities independently of the pulsation period
is by the measurement of the parallax of the stars. Unfortunately, parallaxes
of high significance ($\sigma_{\pi}/\pi < 0.5$) are available for only four of
our sample stars \citep{Pour03}.

In view of this situation, we decided anyway to infer the luminosity from the
pulsation period. In a first step, we used the $\log P - K$ relations of
\citet{Rie10} to determine the absolute $K$ magnitudes. As these relations are
based on LMC stars, we assume that the same relation is applicable to Galactic
AGB stars. This will likely introduce only a small uncertainty in the final
luminosity. For the carbon star \object{BH~Cru}, the relation for sequence 1
(classical Mira sequence) of C-rich stars, and for all other stars sequence 1
of the O-rich stars was used. As the period of \object{RU~Vul} is no longer
clearly detected in the most recent photometric data, we adopted the period of
155\,d that it had before the period decline \citep{TWF08} and forced it to
follow also sequence 1 of the fundamental mode pulsators. This choice is
justified in Sect.~\ref{Discindiv}.
%leaves a larger uncertainty on the absolute magnitude of this star. As the
%brightness difference between sequence 2 and 3 is however not very large, the
%overall picture would not be changed significantly.
A distance modulus to the LMC of 18\fm5 was assumed. To get from absolute
$K$-magnitudes to the absolute bolometric magnitude $M_{\rm{bol}}$, we applied
the BC(K) relations of \citet{KLM10} based on the $J-K$ colour.
Quasi-simultaneous photometry in the $J$ and $K$ bands was collected from
the works of \citet{Catchpole}, \citet{Fouque}, \citet{KH94}, \citet{Whi06}, and
the 2MASS catalogue \citep{2MASS}, transformed to the 2MASS photometric system
with the relations given by \citet{Carpenter}, and averaged to a final value.
For some stars, only the 2MASS data were available, but for most stars this
resulted in a cycle-averaged $J-K$ colour. As in \citet{KLM10}, no correction
for interstellar reddening was applied, because $E(J-K)$ is only a few 0\fm01
for these solar neighbourhood stars and variability and individual circumstellar
extinction have a much larger influence. The BC(K) correction applied ranged
between 2\fm76 and 3\fm11. The mean $J-K$ colour, the absolute $K$ magnitude,
and the resulting bolometric magnitude $M_{\rm{bol}}$ are reported in columns 12,
13, and 14 of Table~\ref{tabres}. The $M_{\rm{bol}}$ derived in this way differ
on average by less than 0\fm08 from what is found by using the relation of
\citet{GB08}.
%\citet[][excluding the SRV RU~Vul]{GB08}.
We estimate a typical uncertainty of $\sim$0\fm16 in the bolometric magnitude
from combining the uncertainties in the $\log P - K$ and BC(K) relations. The
variable period of the stars might introduce an additional uncertainty, which
is hard to quantify.

For the four stars in our sample with a good parallax (\object{R~Aql},
\object{R~Hya}, \object{T~Cep}, and \object{W~Hya}), we can compare the absolute
$K$-magnitude derived from the $\log P - K$ relation with that derived from the
parallax. Only in the case of W~Hya do the values not agree within the error
bars, $M_K$ from the $\log P - K$ relation being brighter by 0\fm36 than the
brightest $M_K$ allowed by the $1\sigma$ uncertainty in the parallax. The
average difference for the other three stars is 0\fm18, in good agreement with
the error estimated for the luminosities adopted here. For \object{R~Nor}, the
uncertainty in the parallax is slightly larger than half the parallax itself.
Nevertheless, a comparison of the $M_K$ magnitudes derived with the two methods
is interesting and suggests that the true parallax is probably near the lower
limit of the $1\sigma$ range, hence the distance is probably near the upper
limit of 800\,pc (see also Sect.~\ref{Discindiv}).
% W Hya falls somewhat below the log P - K relation, if K is determined from the
% parallax (Lebzelter et al. 2005). For the three stars R Aql, R Hya, and T Cep,
% the difference in M_K is on average 0.18mag, in good agreement with the
% estimated uncertainty on M_K derived from the log P - K relations.

\begin{figure}
  \centering
  \includegraphics[width=\linewidth,bb=74 370 549 699]{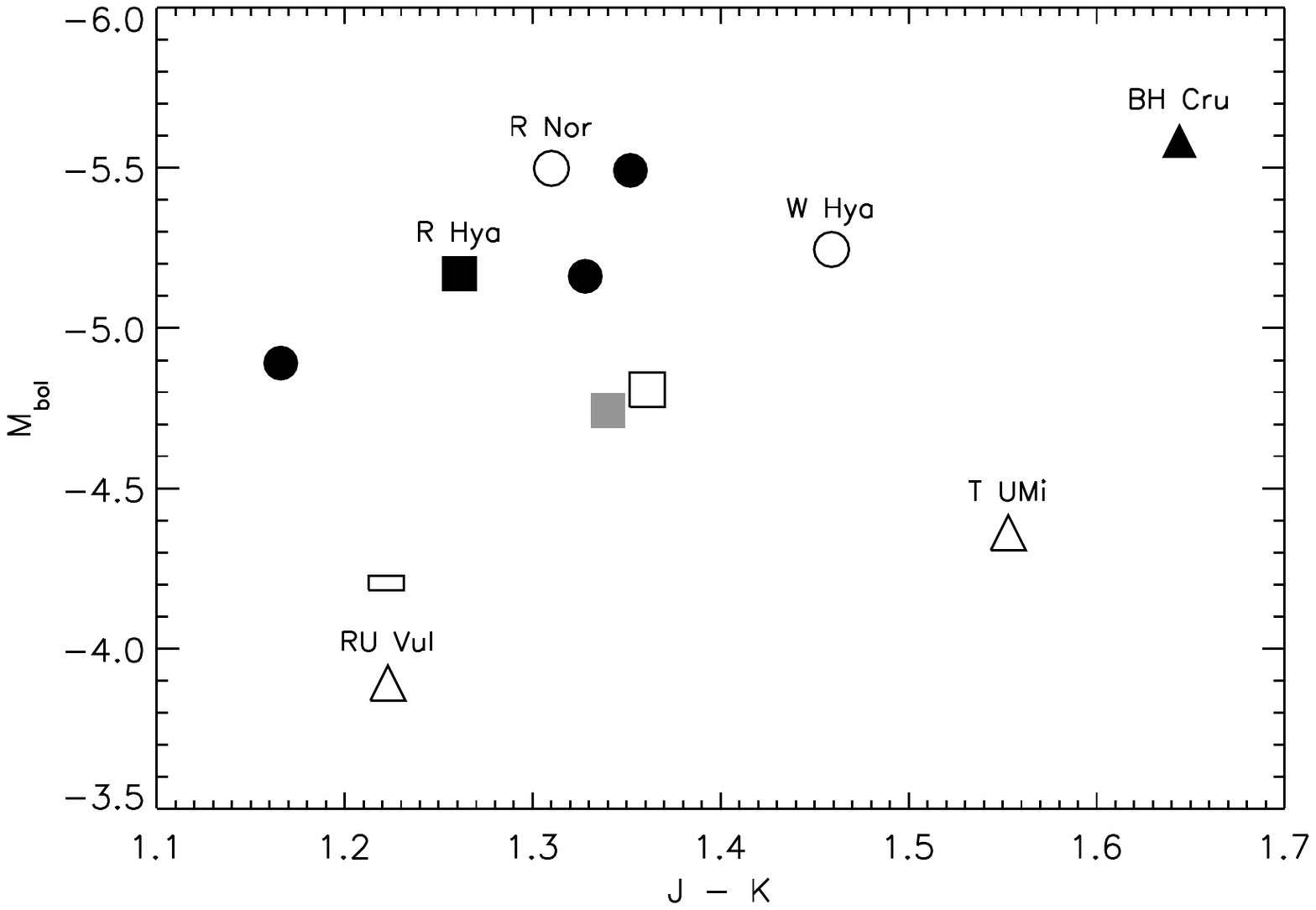}
  \caption{HRD with our sample stars. The symbols are as follows: Triangles
    represent the SCh stars, squares the CCh Miras, circles the MCh Miras,
    and the rectangle represents the constant Mira T~Her. Open symbols are the
    Tc-poor stars, filled symbols Tc-rich. The gray square is W~Dra, for which
    no definite conclusion on its Tc content could be drawn. Stars discussed
    individually in Sect.~\ref{Discindiv} are identified with their names.}
  \label{MbolJK}
\end{figure}

The resulting HRD of our sample stars, using $J-K$ as a proxy for the
temperature, is shown in Fig.~\ref{MbolJK}. As already noted by \citet{Utt07a},
the Tc-rich stars seem to form the upper envelope of the distribution. They
are the brightest stars at any given $J-K$ colour (or temperature). The only
exception here is \object{R~Nor}, a very bright but definitely Tc-poor star.
This is further independent evidence that R~Nor is a massive star with HBB
(see Sect.~\ref{Discindiv}). From its location in the HRD, \object{W~Dra} falls
among the Tc-poor stars, which suggests that it is also Tc-poor.

%\object{T~UMi} is surprisingly red. The only $J$ and $K$ magnitudes that we
%found for this star are from the 2MASS catalogue, so the red colour could be a
%result of a chance observation in the coolest and reddest part of its pulsation
%cycle. However, a comparison of its COBE/DIRBE [1.25] -- [2.2] colour with that
%of other sample stars listed in \citet{Pri10} confirms this red colour. It is
%thus the second reddest sample star after \object{BH~Cru}, which is of SCh
%change class, too. The $K-[12]$ colour, a measure of the mass-loss, is however
%not conspicuously high for T~UMi.
%%Check K -- [12] and K -- Akari colours to see if T~UMi has enhanced mass-loss!
%%Its K -- [12] is not conspicuously red.
%
%The CCh stars occupy a narrow range in $M_{\rm{bol}}$ and $J-K$, but given the
%naturally small sample size it would be very speculative to draw any further
%conclusions from this HRD.

%-------------------------------------------------------------------------------
%-------------------------------------------------------------------------------

\section{Discussion}\label{Disc}

Before we discuss the evolutionary states of the different period change
classes, we first comment on the evolutionary states of a few remarkable
individual objects in the sample. %Finally, the most promising candidates for a
%recent TP are identified. For the other stars, a recent TP cannot be excluded
%by the absence of Tc (only 3DUP is excluded), nevertheless it constrains the
%strength of a possible TP.

\subsection{The evolutionary states of individual objects}\label{Discindiv}
%
%\subsubsection{BH Cru}\label{bhcru}

{\bf\object{BH~Cru}} displays Tc in its spectrum, a clear sign that it is on
the TP-AGB. An s-process enhancement in that star has already been reported by
\citet{AW98}. The pulsation period of BH~Cru has increased considerably since
1970, but has been stable for some 15 years, or even slightly decreased. We
presented in Sections \ref{secCO} and \ref{sectli} evidence that BH~Cru is a
carbon star. It was initially classified as being of SC spectral type
\citep[with C/O essentially unity;][]{CaFe71,KB80}, but its type changed to CS
and it experienced a marked period increase \citep{LE85}. \citet{Zij04}
interpret this spectral change as being due to a decrease in the effective
temperature by some 200\,K, rather than a change in C/O ratio (i.e.\ by a
recent dredge-up of C). The same authors obtained a low-resolution spectrum of
BH~Cru in the range 600 -- 700\,nm and noted that C$_2$ bands are not clearly
present. However, they also stated that \cite{Loidl01} found that C$_2$ bands
are present longward of 700\,nm. We compared the spectrum of BH~Cru to Hermes
spectra of well-known carbon stars and found that \object{RS~Cyg}, a carbon star
of spectral type C8.2, has an almost identical spectrum. The measured C/O
ratios of C-stars are clearly in excess of 1 \citep[e.g.][]{Lam86,Led09}. We
favour the interpretation that the change in spectral type observed in the past
few decades is really due to a recent 3DUP event, and not simply a temperature
decrease. If that were the case, then every normal carbon star could become
either a CS or an SC star simply by {\it increasing} its effective temperature.
%The question whether the spectral change is caused by a change in temperature
%or in C/O ratio will certainly be solved if BH~Cru returns to the period (and
%temperature) that it had at the time of its discovery. If its spectrum remains
%to display strong C$_2$ bands, it would mean that its C/O ratio is now greater
%than 1. Even before such a return, strong constraints on the stellar parameters
%($T_{\rm{eff}}$, C/O) can be obtained with dedicated model atmospheres.
Dedicated modelling of the atmospheres and spectra of cool giant (Mira) stars
with C/O ratio close to unity, including non-equilibrium chemistry, is required
to gain more certainty in this respect. We note that an increase in the C/O
ratio from below to above unity at constant luminosity naturally leads to an
increase in the pulsation period, because of the associated change in chemistry
and opacities. This is suggested by the linear pulsation models employed by
\citet{LW07}, who take into account the effect of the C/O ratio on the
opacities and the atmospheric structure of pulsating AGB stars. This effect
remains to be confirmed with more realistic non-linear pulsation models. If
confirmed, this would mean that the observed period change in \object{BH~Cru}
can actually be caused by the dredge-up of carbon and the thereby changing
opacities, rather than by the associated TP itself.

{\bf\object{RU~Vul}} no longer displays any clearly detectable pulsation, and
shows signs of neither Tc nor Li. We find that it has a very low metallicity of
$[\rm{M}/\rm{H}] \approx -1.5$. This agrees with the finding of \citet{Men01},
who assign it to the extended (thick) disk, which is an old, metal-poor
population. The same authors determine an absolute IRAS~12 magnitude of RU~Vul
(based on the statistical properties of the extended disk sub-sample), from
which we infer a distance of 2070\,pc to that star. If we assume a fundamental
mode pulsation with a period of 155\,d \citep{TWF08}, use relation 1 of
\citet{Rie10} to derive an absolute $K$-magnitude, and combine this value with
the observed 2MASS $K$-magnitude, we arrive at {\it exactly} the same distance.
This provides good confidence that assigning RU~Vul to the fundamental mode
pulsators is the correct choice. \citet{KH92} find that the SRa's are not a
distinct class of variables but a mixture of 'intrinsic' Miras and SRb's.
Before its period and amplitude decline, RU~Vul exhibited a quite regular light
change with a visual amplitude of about $2\fm0$, somewhat below the (arbitrary)
definition limit of Miras, and was classified as SRa. Furthermore, many
short-period, small-amplitude Miras are known to belong to an old, metal-poor
population \citep{Hron}. We conclude that RU~Vul is also representative of this
very same population, but it failed to be classified as Mira because of its too
low pulsation amplitude. Its supposedly high age also suggests that RU~Vul is
of relatively low mass ($\sim 1 M_{\odot}$).

{\bf\object{T~UMi}} is an oxygen-rich, M-type star without any signs of
s-process enrichment. It has thus not undergone a 3DUP event. The change in the
pulsation behaviour of this star is dramatic. Before 1980, the period was fairly
stable at $\sim$315\,d, when it started to decline at a high rate. By mid 2008,
its period had decreased to $\sim$230\,d, when it seems to have switched the
pulsation mode from Mira-like to semi-regular variability (Fig.~\ref{tumilc}).
The present-day period is around 113\,d, with a visual amplitude below the
definition limit of Miras. \citet{White99} already speculate that T~UMi could
drop out of the Mira instability strip in the HRD, which we confirm herewith.
\citet{MF95} suggest that it is in the early helium-shell flash phase where the
period is expected to change most rapidly as the luminosity drops. An
indication for a luminosity drop in T~UMi was found in the detailed light curve
analysis by \citet{SKB03}, who also suggest that the period decrease should
stop in the near future, which is however not (yet) the case. However, from the
lack of Tc (and also Li) it seems that T~UMi is not as highly evolved as other
stars in the present sample, or that it is not massive enough to undergo a 3DUP
event. If a recent TP is the cause of its period change, it is possibly an
early (weak) TP.

%\subsubsection{R Hya}\label{rhya}

{\bf\object{R~Hya}} is the longest-known Mira to have a steadily decreasing
pulsation period. At the time of its discovery around AD 1700, it had a period
of $\sim$500\,d, which then decreased to its present value of $\sim$376\,d, and
might well continue to decline in the future. We find Tc at a low level in this
star, which confirms its evolutionary state on the TP-AGB.
%This finding is contrary to the classification in \citet{UL10}, who use a
%near-maximum spectrum that probably suffered from line weakening.
We thus conclude that the TP scenario for the period change in this star is
likely.

%\subsubsection{R Nor}\label{rnor}

The high Li abundance in {\bf\object{R~Nor}} (Sect.~\ref{sectli}) could be a
result of hot bottom burning (HBB) going on in this star. This would require
the star to have a mass of $\sim 4 M_{\odot}$. \citet{Leb05} suggest that R~Nor
is relatively massive ($3 - 5 M_{\odot}$), because it has a secondary maximum in
its light curve, which is also found in luminous LMC Miras
\citep[see also][]{McS07}. This would be in line with our finding. In the LMC,
the very Li-rich AGB stars are mainly found in the luminosity range
$-6\fm0 \gtrsim M_{\rm{bol}} \gtrsim -7\fm2$ \citep{Smith95}. R~Nor is somewhat
below this limit ($M_{\rm{bol}} = -5\fm50$, Sect.~\ref{Discindiv}), with the
uncertainty that we have to rely on the period to determine its luminosity.
Another indicator of a high mass can be a relatively small distance from the
Galactic plane. The Galactic latitude of R~Nor is 5$^\circ$ and its parallax
$\pi = 2.76\pm1.71$\,mas \citep{Pour03}. Even though the parallax is very
uncertain, this would place R~Nor at a distance from the Galactic plane between
20 and 70\,pc (where the larger distance is the more likely one, see
Setc.~\ref{luminos}). For Miras with periods in the range 300 -- 400\,d, the
exponential scale height is close to 240\,pc \citep{JK92}, hence R~Nor is
relatively close to the Galactic plane. Furthermore, \citet{GH07} find no
s-process enrichment in their sample of Li-rich, massive Galactic AGB stars.
This could be explained by the presence of a massive convective envelope, which
would strongly dilute any material dredged-up from regions close to the core.
In addition, 3DUP could be very inefficient in high mass stars.
%Many of the Li-rich sample stars of \citet{GH07} are also reported to be
%considerably enhanced in rubidium \citep[Rb;][]{GH06}. This is probably a
%signature of the $^{22}$Ne($\alpha$,n)$^{25}$Mg neutron source reaction, which
%is only activated in massive stars ($M > 4 M_{\odot}$).
An additional piece of evidence of a high mass would be an enhanced rubidium
(Rb) abundance, which is synthesised by the $^{22}$Ne($\alpha$,n)$^{25}$Mg
neutron source reaction \citep{GH06}. Unfortunately, our Coralie spectrum of
R~Nor does not cover the Rb line at 780\,nm, so we cannot measure its abundance
in this star. Nevertheless, there is compelling evidence that R~Nor is of high
mass, probably around 4\,$M_{\odot}$.

Whether {\bf\object{W~Hya}} is an SRV or a Mira is not entirely clear. This
question was discussed in detail by \citet{Now10}. From its light variation and
location in the P-L diagram, W~Hya would indeed be classified as a Mira. From
the velocity variations in its atmosphere, on the other hand, it can clearly be
assigned to the group of SRVs. \citet{Leb05} conclude that a high mass might be
the reason for the remarkable velocity behaviour. Although some Li is present
in the atmosphere of W~Hya, the mass of this star is probably too low for
efficient operation of HBB.

\subsection{The evolutionary states of the period change classes}
\label{Discclass}

%\subsubsection{Sudden change}\label{concSCh}

The SCh group in our sample is very heterogeneous: The period of the C- and
Tc-rich Mira \object{BH~Cru} is (or has been) increasing, the O-rich, Tc-poor
stars \object{T~UMi} and \object{RU~Vul} have decreased their periods
considerably.
%In addition, we suggest that this last star is metal-poor
%($[\rm{M}/\rm{H}] \approx -1.5$).
In addition, the derived luminosities (Sect.~\ref{luminos}) differ widely. These
three stars differ quite fundamentally from each other, including in terms of
their evolutionary states. Nevertheless, they show similarly large period change
rates (in absolute terms). If their period change is caused by a TP, the
inhomogeneity of the presence of Tc in the SCh group suggests that a TP with
an associated large period change can happen before a star undergoes a 3DUP
event.

%\subsubsection{Continuous change}\label{concCCh}

From the viewpoint of chemistry and luminosity, the CCh and the MCh groups are
much more homogeneous than the SCh group. All members of these two groups
studied here are late M-type stars. However, both the CCh and the MCh groups
are inhomogeneous in terms of the presence of Tc, even though for
\object{W~Dra} no definite conclusion could be drawn. % While the mildly Tc
%enriched \object{R~Hya} is definitely on the TP-AGB and hence may be regarded
%as a good candidate for a recent TP, \object{R~Aql} is in a preceding phase,
%where TPs are not (yet) strong enough to drive 3DUP.
%If the period change of R~Aql is caused by a TP, then we cannot expect 3DUP to
%happen in connection with this TP, because R~Aql is probably already in the
%declining phase after the maximum luminosity \citep{WZ81}. The period of
%\object{W Dra} is increasing, which means it has not yet reached its maximum
%luminosity during the (possible) TP, and 3DUP could take place at or after this
%maximum (in a few hundred years from now).
%
%\subsubsection{Meandering change}\label{concMCh}
There is some evidence that the MCh group of stars is on average more massive
than the other two groups.
%, {\bf see e.g.\ the discussion about \object{R~Nor} and \object{W~Hya} above}.
The periods in this group are fairly long, between
300 and 500\,d, although this is in some contradiction to \citet{ZB02}, who
note that {\em all} meandering Miras have periods of 400\,d or longer. Four of
five stars in this group have at least some Li, and three display Tc. The
presence of Tc certifies that they are definitely in the TP-AGB phase. We do,
however, not suggest that the period change in the MCh group is connected to a
recent TP, as their periods change back and forth on a much too short
timescale. One of the alternative explanations presented in the introduction
(feedback between pulsation and atmospheric structure or chaotic behaviour)
might hold for them.
% More massive and/or more evolved? Evidence for high mass of R Nor and W Hya,
% the other three stars have Tc, and all have long periods

\section{Summary and conclusions}\label{Concl}

% Umschreiben? Erst: Viele interessante Details ueber einzelne Objekte, BH Cru
% und R Hya as TP Kandidaten, aber wichtigste Schlussfolerung: Keine Korrelation
% zwischen Aenderungstyp und Tc-Vorkommen.
From our search for Tc and other s-process indicators, as well as for Li, in the
spectra of Mira variables with changing pulsation periods, we can draw several
conclusions. In our sample of twelve stars, have identified five that display Tc
absorption lines, indicating that they have undergone a 3DUP event following a
TP. The most important conclusion, however, is that {\it there is no clear
correlation between the presence of Tc and the type of period change}, even in
this limited sample. We suggest that the two Tc-rich, non-meandering Miras
\object{BH~Cru} and \object{R~Hya} are the most likely to have experiences a
recent TP, although their period change rates are quite different. Furthermore,
we presented evidence that BH~Cru changed its spectral type from SC to C, which
indicates that there has been a recent 3DUP that raised its C/O ratio above
unity. There is also accumulating evidence (high Li abundance, high luminosity,
pulsation behaviour, proximity to the Galactic plane) that \object{R~Nor} is a
fairly massive star of $\sim 4 M_{\odot}$. On the basis of an analysis of AAVSO
data of \object{T~UMi}, we report that this star probably switched its
pulsation mode from Mira-like to SRV very recently. Finally, we suggest that
\object{RU~Vul} is a representative of the old, metal-poor thick disk of a
consequently low mass.
%This, however, requires some explanation
%of their very different period change rates. As suggested in Sect.~\ref{bhcru},
%this could be reconciled if the now sharp period increase of BH~Cru is caused
%by a recent 3DUP that increased the C/O ratio above unity, and the envelope is
%now adapting to the opacities of the C-rich atmosphere. It would be worthwhile
%to investigate the evolution of the pulsation period and mode of a Mira with a
%non-linear pulsation model, if the opacities of the atmosphere are relatively
%quickly changed from O- to C-rich chemistry.

Continuous photometric monitoring of Mira variables, in particularly those with
large period changes, is mandatory to gain further insight into the AGB phase
of stellar evolution. The archives of amateur observations such as those at
AAVSO are an invaluable resource of data for this study. In parallel to
photometric monitoring, we also recommend {\em spectroscopic} monitoring of the
stars with changing pulsation period, albeit with sparser sampling of a few
years. This will be of high importance to check whether the period change is
also accompanied by a spectral change.
% (as is clearly the case for \object{BH~Cru}).

In addition to the prominent and well-studied objects presented here, there are
more stars with pronounced period changes that deserve more detailed scrutiny
in the future. \object{LX~Cyg} \citep{Zij04} is a particularly interesting
example, which is classified as SC spectral type, just as \object{BH~Cru} was
before its spectrum started to change. Assuming that the s-process enrichment
in LX~Cyg is intrinsic, this star would similarly be placed on the TP-AGB. A
period change caused by a recent TP seems plausible.
%, although \citet{Zij04} suggest that it could instead be connected to a
%feedback effect between changing temperature, opacities, and pulsation
%behaviour of the envelope.
Furthermore, \object{TT~Cen} \citep{White99}, \object{S~Sex} \citep{MBJV00},
as well as the remarkable objects mentioned in \citet[][\object{R~Cen},
\object{Z~Tau}, \object{Y~Per}]{TMW05} deserve a closer look with
high-resolution spectroscopy, which we have to postpone to a future project.

\begin{acknowledgements}

  The Hermes project is a collaboration between the K.U.\ Leuven, the
  Universit\'{e} Libre de Bruxelles, and the Royal Observatory of Belgium with
  contributions from the Observatoire de Gen\`{e}ve (Switzerland) and the
  Th\"uringer Landessternwarte Tautenburg (Germany). Hermes is funded by the
  Fund for Scientific Research of Flanders (FWO) under the grant G.0472.04, from
  the Research Council of K.U.\ Leuven under grant GST-B4443, from the Fonds
  National de la Recherche Scientifique under contracts IISN4.4506.05 and FRFC
  2.4533.09, and financial support from Lotto (2004) assigned to the Royal
  Observatory of Belgium. The authors wish to acknowledge the Geneva Observatory
  and its staff for the generous time allocation on the Swiss Leonhard Euler
  telescope. We wish to thank the Hermes and Coralie observers who carried out
  the observations used for this research: P.\ Beck, P.\ Degroote, C.\ Gielen,
  R.\ \O stensen, W.\ Pessemier, and T.\ Verhoelst. %include them as co-authors?
  SU acknowledges support from the Fund for Scientific Research of Flanders
  (FWO) under grant number G.0470.07, and from the Austrian Science Fund (FWF)
  under project number P~22911-N16, FK from P~18939-N16. This publication makes
  use of data products from the Two Micron All Sky Survey, which is a joint
  project of the University of Massachusetts and the Infrared Processing and
  Analysis Center/California Institute of Technology, funded by the National
  Aeronautics and Space Administration and the National Science Foundation. We
  acknowledge with thanks the variable star observations from the AAVSO
  International Database contributed by observers worldwide and used in this
  research. We thank the anonymous referee for constructive comments, which
  helped to improve the paper.
  %mail them at aavso@aavso.org

\end{acknowledgements}

\end{document}